\documentclass[aps,prb,twocolumn,showpacs,amsmath,amssymb]{revtex4-1}
\usepackage{txfonts}
\usepackage{graphicx}
\usepackage{cases}
\usepackage{breqn}
\usepackage{amsmath}
\usepackage{dcolumn}
\usepackage{bm}
\usepackage{multirow}
\usepackage{float}
\usepackage{tabularx}
\usepackage{color}
\usepackage[dvipsnames]{xcolor}
\usepackage{hyperref}
\usepackage{graphicx}
\usepackage{subcaption}
\usepackage[export]{adjustbox}

\def\eV{\,\textrm{eV}}

\def\Mott{{\textrm{Mott}}}
\def\cond{{\textrm{cond.}}}
\def\min{{\textrm{min}}}
\def\evan{{\textrm{evan}}}
\def\tot{{\textrm{tot}}}

\begin{document} 
 
\title{DFT+DMFT calculations of the complex band and tunneling behavior for the transition metal monoxides MnO, FeO, CoO and NiO} 

\author{Long Zhang$^{1}$ } 
\author{Peter Staar$^{2}$ } 
 \altaffiliation[current address: ]{Cognitive Computing and Industry Solutions, IBM Research, Saumerstrasse 4, 8803 Rueschlikon, Switzerland \\}
\author{Anton Kozhevnikov$^{3}$ }
\author{Yun-Peng Wang$^{1}$ }
\author{Jonathan Trinastic$^{1}$ }
\author{Thomas Schulthess$^{2,3}$ }
\author{Hai-Ping Cheng$^{1}$ }
  \email{Contact Email: cheng@qtp.ufl.edu; Tel: 352-392-6256 \\ }

\affiliation{$^{1}$ Department of Physics and The Quantum Theory Project$,$  University of Florida$,$  Gainesville FL 32611$,$ USA}
\affiliation{$^{2}$ Institute for Theoretical Physics$,$ ETH Zurich$,$ 8093 Zurich$,$ Switzerland}
\affiliation{$^{3}$ Swiss National Supercomputing Center$,$ ETH Zurich$,$ 6900 Lugano$,$ Switzerland}


\begin{abstract}

We report complex band structure (CBS) calculations for the four late transition metal monoxides, MnO, FeO, CoO and NiO, in their paramagnetic phase. The CBS is obtained from density functional theory plus dynamical mean field theory (DMFT) calculations to take into account correlation effects. The so-called $\beta$ parameters, governing the exponential decay of the transmission probability in the non-resonant tunneling regime of these oxides, are extracted from the CBS. Different model constructions are examined in the DMFT part of the calculation. The calculated $\beta$ parameters provide theoretical estimation for the decay length in the evanescent channel, which would be useful for tunnel junction applications of these materials. 


\end{abstract}
\pacs{}


\maketitle

\section{ \label{sec:1} INTRODUCTION } 

Motivated by the application of transition metal oxides (TMO) in modern electronics, the charge transport through TMO nano-junctions has been extensively investigated in the past ~20 years, theoretically and experimentally. A large amount of literature focuses on the non-resonant tunneling experiments in which the tunneling current decays exponentially, $I=I_{0}\cdot exp(-\beta L)$, as the length of the tunnel junction ($L$) increases. Although the $\beta$ parameter depends on the inter-facial properties between the junction and the metallic electrodes, it is mainly determined by the electronic properties of the junction material itself. Since many TMOs are strongly correlated electronic systems, calculation of $\beta$ from first principles with the inclusion of electron correlation would be necessary and important for understanding TMO nano-junctions. In this work we report \textit{ab initio} calculations of $\beta$, based on DFT and single-site DMFT, for the late TMO monoxides. 

Existing studies have shown the $\beta$ parameter is related to a material's band gap, the hopping parameter $t$ of the insulating material, and the alignment of the Fermi level in the metal electrodes with the band gap of the insulating junction \cite{PhysRevB.65.245105,PhysRevB.85.235105}. One way to calculate $\beta$ from first principle is to evaluate the complex band structure (CBS)\cite{0370-1328-89-2-327} rather than the ordinary real band structure (RBS). Complex band structure is the energy eigenvalues defined for complex values of $\vec{k}$. The wavefunction of a crystal structure has the well-known Bloch form of $\psi=\psi_{0}e^{i\vec{k}\cdot\vec{r}}$, where $\vec{k}$ is real. In fact, only the solutions of Schr{\"o}dinger's equation with real wave vector are considered, and wavefunctions having complex wave vectors are often not considered because they would grow exponentially in some direction which is physically unreasonable for periodic systems. However, when dealing with surfaces and interfaces of semiconductors, i.e. finite or non-periodic solids, solutions with complex wave vectors have physical meaning for energies within the band gap. They represent the states exponentially decaying into the semiconductor, also called evanescent interface-induced gap states. If $\vec{k}$ becomes a complex variable $\vec{k} = \vec{k}_{Re} + i\vec{k}_{Im}$, then the wavefunction can be written as $\psi=\psi_{0}e^{i\vec{k}\cdot\vec{r}} = (\psi_{0}e^{-\vec{k}_{Im}\cdot\vec{r}})e^{i\vec{k}_{Re}\cdot\vec{r}}$, yielding an exponential decay factor to the amplitude of the wavefunction. The decay is for the direction $\vec{k}_{Re}$, and corresponds to the electron non-resonant tunneling current decay $I=I_{0}\cdot exp(-\beta L)$ in that direction. The $\beta$ parameter is associated with the imaginary part, $\vec{k}_{Im}$, via the relation $\beta = 2|\vec{k}_{Im}|$. The energy bands are generalized to be defined on contours in the complex plane of $\vec{k}$. A very useful feature of CBS is that one can directly read $\vec{k}_{Im}$ thus $\beta$ from the band structure without additional calculations. By picking an arbitrary energy of the wavefunction within the gap, one can trace sideways to the nearest complex band at that energy level and trace down to the corresponding $\vec{k}_{Im}$. In many cases, it is sufficient to apply the CBS approach with the standard Kohn-Sham (KS) density functional theory (DFT). However, it could yield wrong results for materials in which electron correlation plays an important role. The self-energy due to correlation must be considered in such cases. There have been CBS studies based on beyond-DFT calculations. For example, the GW approximation and hybrid density functionals had been used to calculate $\beta$ of simple organic molecules and yielded better agreement with experimentally known values \cite{PhysRevB.65.245105,PhysRevB.85.235105}. In this study, we analyze the CBS and calculate the $\beta$ of Mott insulating materials, using NiO, CoO, FeO and MnO as examples. The correlation effect is taken into account by carrying out DFT plus DMFT calculation and $\beta$ is evaluated from the DMFT-corrected band structure. 


The series of 3$d$ transition metal monoxides with rock salt structure present very rich physical properties. Early in the series, TiO and VO are metallic materials, whereas later members, $e.g.$ NiO, CoO, FeO and MnO, show clear insulating properties and antiferromagnetic (AFM) ordering below the Neel temperature $T_{N}$. At room temperature, NiO is in AFM phase and the other three are in paramagnetic (PM) phase ($T_{N} = ~525K, ~290K, ~198K, ~120K$ for NiO, CoO, FeO and MnO, respectively) \cite{PhysRevB.70.195121,PhysRevB.65.125111,PhysRevB.67.184420,article_Neel_T_FeO}. Because most of the tunnel junction applications are operated at room temperature or even lower temperatures, carrying out the calculation of wavefunction decay rate in the PM phase of these materials needs to be justified, at least for NiO. This is supported by several experiments as well as developments in the calculation side. It is well known that DFT in the local density approximation (LDA) fails to provide a band gap for these materials. Better matching between the experimental data and the calculated band structures was reached first by using the spin-resolved version of LDA, the local spin density approximation (LSDA) \cite{doi:10.1139/p80-159,0022-3719-5-13-012}. Independent angle resolved photo-emission experiments \cite{PhysRevB.44.3604,TJERNBERG19951215} in the 1990s studied the valence band property of bulk NiO below 525K. The experiments demonstrated that the valence band structure from a LSDA calculation with AFM ordering agreed with the experimental data better than LDA. However the band structure close to Fermi energy was still very different from the measured data and the calculated band gap was too small ($< 1 \eV$ vs. $>4 \eV$). It was later experimentally observed that there was actually no change in the photo-emission spectra features during the AFM $\rightarrow$ PM transition of NiO \cite{PhysRevB.54.10245}, which contradicts theories that the band gap of NiO is mainly due to AFM ordering. On the calculation side, the development of the LDA+U method provided a much improved band gap of $ 3.4 \eV $ for NiO \cite{PhysRevB.60.10763,book_TMO}, and the LSDA+U \cite{0022-3719-5-13-012,book_DensityFunctionalTheoryOfAtomsAndMolecules} calculations also had success in describing the electronic structure of 3$d$ metal monoxides. These facts suggest that the band gap of NiO is mainly due to electronic correlation rather than AFM ordering. The decay rates of evanescent channel calculated in the PM phase should not be significantly different than in the magnetic ordered phase, because the $\beta$ is mainly related to the materials' band gap and the band gap is not significantly affected from the AFM $\rightarrow$ PM transition. For CoO, FeO and MnO, we do not find similar experiments studying whether the photo-emission spectra features change in AFM $\rightarrow$ PM transition. All three are in PM phase at room temperature. 


The calculations presented here are carried out in a straightforward way. The four materials' ground state band structures are first calculated using the full potential linearized augmented planewave (FP-LAPW) method. The obtained band structures are then used to construct effective Hamiltonian in Wannier orbital basis, and also used to compute Coulomb interaction matrices using the cRPA method \cite{PhysRevB_frequ_depend_U}. With the Hamiltonian and the $U$ matrices, we perform DMFT calculations to get the $k$-resolved spectral functions and analyze the band gaps. Using the DMFT self-energy, we construct the full Green's function and use it to calculate the complex band structures (CBS) and extract the decay rate.

$\mathit{Outline:}$ The remainder of the article is organized as follows. Section \textcolor{blue}{II} introduces the calculation methods, including the DFT plus DMFT scheme and the way we obtained the $\beta$ parameter from CBS. The essential step of computing the Coulomb interaction $U$ matrices are grouped in Appendix A. The resulting spectral function, as well as the $k$-dependence of the $\beta$ parameter, are described in Section \textcolor{blue}{III}. Section \textcolor{blue}{IV} provides the conclusion. 


\section{ \label{sec:2} METHODS }

The CBS can be calculated using either wavefunctions or Green's functions. We used the Green's function approach because it's consistent with the DMFT formalism. The self-energy from DFT+DMFT is used to construct the full Green's function, which is then used to evaluate the CBS and $\beta$. We will first describe our DFT and DMFT calculation, then explain how we calculate CBS from Green's function, and how we find $\beta$ from CBS. 

The four monoxides, especially NiO, have been extensively studied in the DMFT community and used as benchmark material for novel computational methods \cite{Korotin2008,M.Karolak_double_counting_NiO,PStaar_PRB_continuous_pole}. The existing DFT+DMFT calculations of NiO were not done in the exactly same way. One difference in our calculation is the use of cRPA method to calculate the $U$ matrices in the same Wannier orbital basis used for the Hamiltonian construction. Thus the hopping and interaction parameters of the effective Hubbard model are consistently built from the same DFT ground state. 

\subsection{ DFT calculation }

\begin{figure}[H]
  \includegraphics[width=1.0\columnwidth]{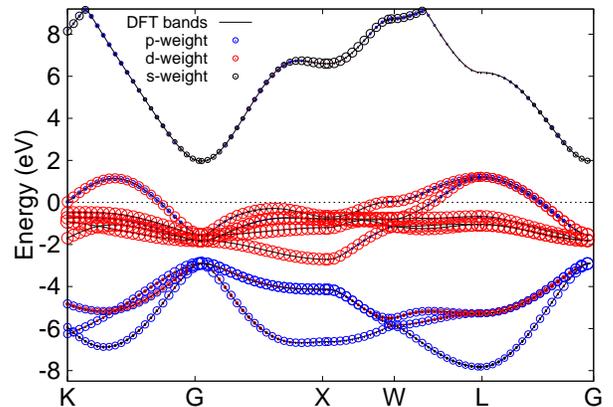}
  \caption
  {Non-magnetic NiO ground state band structure (solid lines), and band characters (open circles) calculated by projecting Bloch states onto atomic orbital states. The radius of the open circles are proportional to the weight of the atomic states. Fermi level is at zero.} 
  \label{fig:NiO_fat_bands}
\end{figure}

Our DFT calculation is done using the FP-LAPW method, as implemented in a modified version of the ELK code \cite{Anton_IEEE_paper_2010}. The ground state is calculated within the generalized gradient approximation (GGA) using the PBE functional. The muffin tin sphere radii are, for example, 2.02 $a_{0}$ and 1.72 $a_{0}$, for Ni and O, respectively. The experimental values of lattice constants are used \cite{PhysRevB.70.195121,PhysRevB.65.125111,PhysRevB.67.184420,article_Neel_T_FeO}. A dense k-point grid of $16$x$16$x$16$ was used to perform Brillouin zone integration. Figure \ref{fig:NiO_fat_bands} displays the ground state band structure of NiO and orbital characters (the amount of overlapping between Bloch states and atomic orbital states). 

\begin{figure}[H]
    \centering
    \begin{subfigure}{.23\columnwidth}
        \includegraphics[scale=0.25]{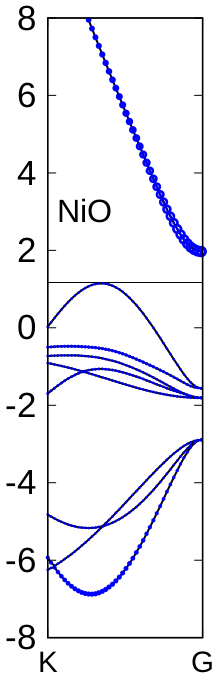}
        \caption{}
    \end{subfigure}
    \begin{subfigure}{.23\columnwidth}
        \includegraphics[scale=0.25]{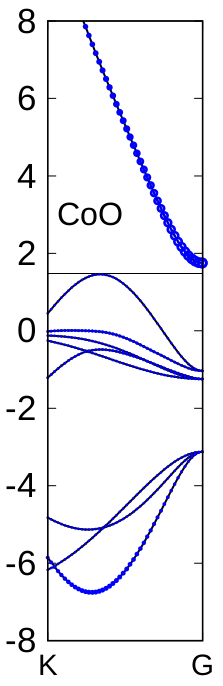}
        \caption{}
    \end{subfigure}
    \begin{subfigure}{.23\columnwidth}
        \includegraphics[scale=0.25]{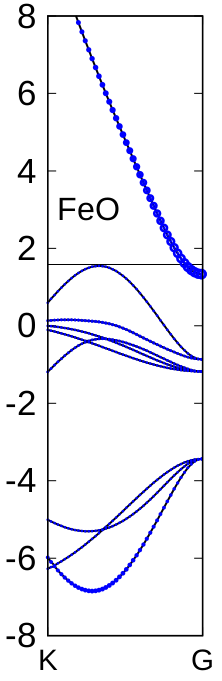}
        \caption{}
    \end{subfigure}
    \begin{subfigure}{.23\columnwidth}
        \includegraphics[scale=0.25]{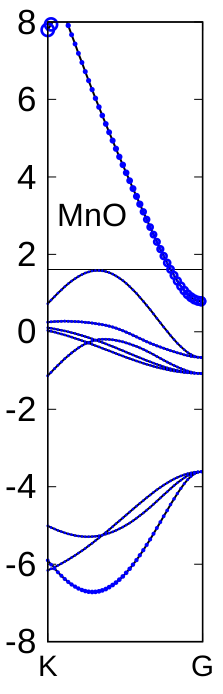}
        \caption{}
    \end{subfigure}
    \caption{Non-magnetic ground state band structures (solid curves) of NiO, CoO, FeO and MnO. And the TM 4$s$ orbital character only (open circles). Horizontal solid lines are placed at the maximum of $d$-like bands, to make the separation or overlap of $s$-like and $d$-like bands clearly seen. Fermi level is always at zero.}
    \label{fig:4s_at_Gamma_point}
\end{figure}

\noindent It is clearly seen that there are five $d$-like bands around the Fermi level, representing the partially filled $d$ states of the transition metal atom and giving the material a metallic state. Below them in the  $ [-8.0,-2.0] \eV $ range are three bands showing $p$-orbital character. Above the $d$-like bands, in the $ [+2.0,+8.0] \eV $ range, there is a single band of transition metal (TM) $4s$ character. It is a common feature of the four materials that the $d$-like bands and $p$-like bands are separated by a small gap. The group of $p$-like and $d$-like bands are isolated from lower bands, but are very close to the $s$-like band at the $\Gamma$ point. Within the GGA-PBE calculation, as shown in Figure \ref{fig:4s_at_Gamma_point}, we find that the $s$-like band is slightly gapped from the $d$-like bands in the cases of NiO and CoO but is overlapping in energy with the $d$-like bands for FeO and MnO. Though not shown, we also find that, when using the LDA functional, the $s$-like band has overall more overlap with the $d$-like bands for these four materials. 

\begin{figure}[H]
  \includegraphics[width=0.9\columnwidth]{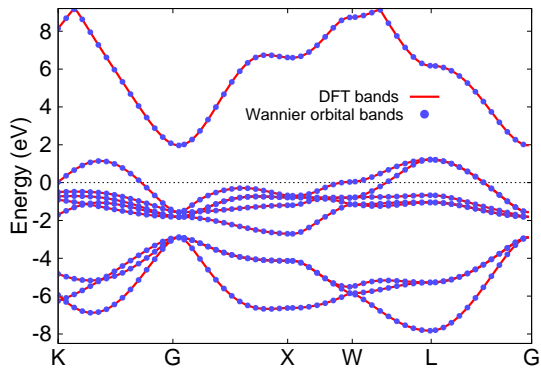}
  \caption
  {DFT band structure of NiO (solid line), and reconstructed bands (dots) in symmetry-preserving Wannier orbital basis, which are identical to DFT bands by construction. Fermi level is at zero.} 
  \label{fig:wannier_vs_dft}
\end{figure}

\noindent The orbitals' characters in Fig.\ref{fig:NiO_fat_bands} and Fig.\ref{fig:4s_at_Gamma_point} display a clear $d$-$p$ mixing in these materials, which motivates our model construction explained in next section. The $s$ orbital weight is well located in the singe band in the $ [+2.0,+8.0] \eV $ range. We do not observe significant mixing between $s$ and the group of $p$ and $d$. We will keep the $s$-like band in the analysis because it had been demonstrated in existing DFT+DMFT studies \cite{PhysRevB.92.085142,Nekrasov_JETP2013} that the TM $s$-like band has significant contribution to the photo emission spectrum of these monoxides. In addition, as we will see in later sections, the extension of the $s$-like band in the complex domain goes across the gap region; thus it should be included for a correct determination of Fermi level pinning position. So, in order to construct localized orbital basis for all later calculations, we downfold the Bloch bands in the energy window $[-8.0,+8.0] \eV $ to the symmetry-preserving Wannier orbital basis that includes the TM $d$-like bands, the TM $s$-like band and the Oxygen $p$-like bands. The reconstructed bands of NiO are shown in Fig. \ref{fig:wannier_vs_dft}.

\subsection{ DMFT calculation }

Dynamical mean field theory is one successful way to more accurately capture the electronic correlation effect and remedy the failure of DFT. Application of DMFT to TMOs originated from the work of Peierls and Mott \cite{Mott_paper_1937, Mott_paper_1949}. Usually an appropriate correlation subspace was identified as those electron states in the partially-filled, transition-metal $d$ shell, and was associated with interactions including the on-site intra-$d$ and inter-$d$ interactions. During the past two decades, the DMFT method \cite{RevModPhys_1996} has been developed for the low energy effective Hubbard model constructed for real materials' $d$-like or $f$-like bands. The widely-adopted numerical scheme involves selecting DFT bands near the Fermi energy as the correlation subspace and fitting them to a tight-binding model using the downfolding technique applied to localized orbitals, such as Wannier orbitals \cite{PRB_muffin_tin_orbital,PhysRevB_MLWF_1997,RevModPhys_MLWF_2012}. For each $\vec{k}$ point, the Bloch Hamiltonian is downfolded to the Wannier orbital basis \cite{Anton_IEEE_paper_2010}, $H^{Wann}(\vec{k})$. Through the Fourier transformation, $H^{Wann}(\vec{k})$ serves as single particle hopping $t_{ij}$ in the first term of Eq.(1) below. This Hamiltonian contains contributions from the effective potential of the DFT calculation that also creates a double-counting issue, which is explicitly accounted for by a correction within DMFT. 


The multi-orbital Hubbard model Hamiltonian with on-site Coulomb interaction can be expressed within the second quantization framework as \cite{Hubbard_paper_1963} 
\begin{multline} 
\hat{H}_{\textrm{Hubbard}} = \hat{H}_{\textrm{Kinetic}} + \hat{H}_{\textrm{Coulomb}} \\
= \sum_{i,j,\mu,\nu} t_{ij, \mu\nu}^{dp} \, \hat{c}_{i}^{+} \, \hat{c}_{j} + \frac{1}{2} \sum_{i,\alpha,\beta,\gamma,\delta} U_{i,\alpha\beta\gamma\delta}^{d} \, \hat{c}_{i,\alpha}^{+} \, \hat{c}_{i,\beta}^{+} \, \hat{c}_{i,\gamma} \, \hat{c}_{i,\delta}
\end{multline}
Here the indices $i,j$ are site indices, and $\mu$,$\nu$ are orbital indices including spin for all orbitals within the correlation subspace. The indices \{$\alpha,\beta,\gamma,\delta$\} are a subset of \{$\mu$,$\nu$\} to indicate those orbitals associated with on-site Coulomb interactions. For the four monoxides in this study, due to the $d$-$p$ mixing mentioned in Sec.II.A, the correlation subspace includes both $d$ and $p$ orbitals and the subset \{$\alpha,\beta,\gamma,\delta$\} of interacting orbitals is limited to $d$ only. The Coulomb interaction tensor $U_{i, \alpha\beta\gamma\delta}^{d}$ can be computed from first-principle or sometimes used as empirical parameters of the model. With the hopping and Coulomb interaction parameters at hand, the DMFT method iteratively solves the model by mapping it to an effective Anderson single-impurity model (ASIM). The impurity Green's function is often expressed in the path integral formulation, with integration over Grassmann fields of second quantization creation and annihilation operators, $\hat{c}^{+}$ and $\hat{c}$, 

\begin{equation}
G_{imp}(i_{1},\tau_{1};i_{2},\tau_{2})=-\frac{\int{D[\hat{c}^{+}]D[\hat{c}]}e^{-S[\hat{c}^{+},\hat{c}]}\left\{\hat{c}(\tau_{1}) \hat{c}^{+}(\tau_{2})\right\}} {\int{D[\hat{c}^{+}]D[\hat{c}]}e^{-S[\hat{c}^{+},\hat{c}]}}
\end{equation}
In Eq.(2), $D[.]$ is the standard integration measure. $S[\hat{c}^{+},\hat{c}]$ is the effective action as defined in Eq.(3) below for the impurity. 
\begin{multline}
S[\hat{c}^{+},\hat{c}]=-\int_{0}^{\beta}d\tau\int_{0}^{\beta}d\tau\ensuremath{'}\sum_{i,j}\hat{c}_{i}^{+}(\tau) \mathcal{G}_{0,ij}^{-1}(\tau-\tau\ensuremath{'})) \hat{c}_{j}(\tau\ensuremath{'}) + \\
 \int_{0}^{\beta}d\tau \hat{H}_{Coulomb}(\hat{c}^{+},\hat{c})
\end{multline}
In Equations (2) and (3), $i$ and $\tau$ are the site index and imaginary time. $\mathcal{G}_{0,ij}$ is the bare propagator, which is also called the \textit{bath} Green's function. It plays a similar role as the Weiss field in classical mean-field theory. Specifically, it describes an effective field coupled to the impurity that contains all non-local information of the underlying lattice, and the lattice is considered as a reservoir of non-interacting electrons. The difference from the classical Weiss field arises in its time dependence, which accounts for local dynamics. We refer readers to Ref.\cite{book_CondensedMatterFiledTheory} for explicit definitions of $D[.]$ and $S[\hat{c}^{+},\hat{c}]$. 

Given the effective action, there exists several well established numerical methods to solve for the impurity's Green's function. The family of quantum Monte Carlo (QMC) solvers are widely accepted and numerically exact if the simulation time is sufficiently long. We refer readers to Ref.\cite{RevModPhys.83.349} for technical details about general QMC impurity solver. In this work, we are using the continuous time hybridization expansion (CT-HYB) QMC solver implemented in the DCA++ code \cite{PStaar_PRB_continuous_sigma_dca,DCA_plusplus_code}. The solver adopts the segment picture \cite{RevModPhys.83.349} to take into account density-density interactions. The coupling to the \textit{bath} is diagonal only in orbital space. Our calculations are performed at inverse temperature $1$/$kT=20$. The number of Monte Carlo sweeps in the QMC calculation is $10^{6}$ in each solver run. The continuous-pole-expansion method \cite{PStaar_PRB_continuous_pole} is used for obtaining the self-energy and impurity Green's function in real frequency domain.

For material-specific calculations, the \textit{lattice} Green's function is constructed, within the correlation subspace, from the downfolded Hamiltonian: 
\begin{equation} 
G(i\omega_{n}) = \frac{1}{N_{k}}\sum_{\vec{k}}\frac{1}{(i\omega_{n}+\mu)-H^{Wann}(\vec{k})-\Sigma(i\omega_{n})} 
\end{equation} 
where $N_{k}$ is the number of $\vec{k}$ points, $\omega_{n}$ is the Matsubara frequency and $\mu$ the chemical potential. The self-energy $\Sigma(i\omega_{n})$ is supplied with an initial guess, then updated in each of the DMFT iteration. One uses the Dyson's equation in each iteration to derive the \textit{bath} Green's function and the effective impurity problem numerically, $i.e.$ $\mathcal{G}_{0}^{-1}(i\omega_{n}) = G(i\omega_{n})^{-1} + \Sigma(i\omega_{n})$, and solve the impurity problem. For the late TM monoxides with strong $d$-$p$ mixing, we include five $d$ and three $p$ orbitals in the correlation window, while limiting interactions to $d$ orbitals. Thus the Hamiltonian $H^{Wann}(\vec{k})$ and lattice Green's function $G(i\omega_{n})$ in Eq.~(4) are eight-dimensional. $\Sigma(i\omega_{n})$ is always five-dimensional, and is added to the $d$-block of $H^{Wann}(\vec{k})$. When constructing the bath $\mathcal{G}_{0}^{-1}(i\omega_{n})$ for interacting orbitals, we use the $d$-block of $G(i\omega_{n})$ together with $\Sigma(i\omega_{n})$ in Dyson's equation.


The value of interaction parameters in $\hat{H}_{Coulomb}(\hat{c}^{+},\hat{c})$ are calculated using \textit{ab initio} methods from the materials' DFT ground states. The constrained Random Phase Approximation (cRPA) method, as explained in details in Appendix A, is adopted for this step. The important step in cRPA is to choose a screening window, within which the particle-hole polarization are excluded. The $d$-$p$ mixing gives some arbitrariness here because one cannot find a window of bands that includes exactly all $d$-weight and exclude all $p$-weight. There are naturally two choices: excluding both $d$-$like$ and $p$-$like$ bands in [-8.0,+2.0] eV which is often called the $dp$ model; excluding only the five $d$-$like$ bands which is called the $d$-$dp$ model. We have calculated the on-site Coulomb interactions of the five Wannier $d$ orbitals for the two models. The results are discussed in Appendix A. The later complex band analysis is build on the $d$-$dp$ model only. We present the cRPA results of both models in this work for the partial purpose of benchmarking the current cRPA implementation.


The DFT+DMFT calculation scheme and its variants have been widely used in the past two decades to study TMOs that have a pronounced correlation effect. It is worth briefly reviewing the existing studies and pointing out the differences and limitations in the current work. One of the earliest applications using DFT+DMFT for real materials was a study of NiO \cite{PhysRevB.74.195114}, where a realistic gap and the near-gap spectra were obtained for a correlation subspace that contains $d$ orbitals only. Shortly after, the oxygen $p$ states of NiO were included \cite{PhysRevB.75.165115,PhysRevLett.99.156404}, in a way similar to that described in this section, to provide fuller description of the valence-band spectrum. It was found in these studies that doping holes leads to the filling of the correlation gap and a significant transfer of the $d$ spectral weight. In these studies the low-energy Zhang-Rice bands were also obtained for the first time. Besides the paramagnetic state, magnetic state properties of NiO were also investigated within the framework of DFT+DMFT \cite{PhysRevB.77.195124,PhysRevB.85.235136}, where the Iterated Perturbation Theory (IPT) solver and the numerical exact diagonalization (ED) solver were used to solve the impurity problem. NiO has also been actively used as a benchmark material for DFT+DMFT method development, \textit{e.g.} new methods related to the double counting correction \cite{M.Karolak_double_counting_NiO,Nekrasov_JETP2012,Nekrasov_JETP2013} and analytical continuation \cite{PStaar_PRB_continuous_pole}. The other members of the late TM monoxides, CoO, FeO and MnO, together with NiO have been studied within the DFT+DMFT scheme in a systematic investigation of the band gaps of these materials \cite{Nekrasov_JETP2013}, investigation of the fundamental quantum entanglement of indistinguishable particles \cite{PhysRevLett.109.186401}, and as benchmarking materials in method developments for determining the Coulomb correlation strength \cite{U_from_MLWF_cRPA_NiOCoOFeOMnO_2013,PhysRevLett.108.087004}. 

Besides studies under ambient conditions, there are a significant number of first-principle calculations focusing on the beyond-equilibrium properties of the late-TM monoxides, particularly the changes of electronic structure related to high pressure and lattice distortions. It was first reported \cite{Nat.Mater.7.198} that MnO experiences a simultaneous moment collapse, volume collapse, and metallization transition under a pressure of about $100$ GPa. Upon compression of $60-70$ GPa, the B1 structure of FeO has a spin-state transition accompanied by an orbital-selective Mott metal-insulator transition \cite{PhysRevB.82.195101,PhysRevB.92.085142}, in good agreement with the experimental result \cite{PhysRevLett.108.026403}. A pressure-driven orbital selective insulator-to-metal transition is also observed in CoO \cite{PhysRevB.85.245110,Dyachenko2012}. Similar to what is seen in FeO, the $t_{2g}$ orbitals of Co become metallic first at about $60$ GPa, and the $e_{g}$ orbitals remain insulating until the much higher pressure of about $170$ GPa. It is found that the transition to fully metallic state is driven by a high-spin to low-spin transition of the Co$^{2+}$ ions. A systematic study of all four TM monoxides under high pressure \cite{PhysRevB.94.155135} reveals a remarkably high pressure of $430$ GPa for the insulator-metal transition in NiO, which is well out of the range $170-40$ GPa of MnO--CoO. 

The full charge-density self consistent (CSC) DFT+DMFT scheme is often used in TM oxide calculations under high pressure because the charge density is subject to change with lattice distortion. In the CSC scheme \cite{GRANAS2012295,PhysRevB.90.235103,PhysRevB.94.155131}, the DMFT iteration described in this section is nested in an outer iteration of the charge density. The many-body effect within the correlation subspace is self-consistently included in the entire system. It has been demonstrated that the CSC scheme is necessary in studying the metal-insulator transition of V$_{2}$O$_{3}$ \cite{PhysRevB.91.195115}, where a strong enhancement of the $a_{1g}-e^{\pi}_{g}$ crystal-field splitting causes a substantial re-distribution of charge density and thereby influences the lattice structure due to electron-lattice coupling. In a recent study of pressure-induced insulator-metal transition in Fe$_{2}$O$_{3}$ \cite{PhysRevX.8.031059}, a site-selective redistribution of the Fe $3d$ charges between the $t_{2g}$ and $e_{g}$ orbitals associated with spin state transition was captured within a CSC DFT+DMFT calculation. The CSC scheme might be important for the current study mainly because the $4s$-like band enters the correlation window at the $\Gamma$-point. If the $4s$-like band is significantly shifted in a CSC DFT+DMFT calculation then its complex extension would be shifted too and affect the complex band structure within the Mott gap. Indeed, in existing CSC DFT+DMFT studies of late TM monoxides under pressure \cite{PhysRevB.92.085142,PhysRevB.94.155135}, the $4s$-like band is significantly lowered (by about $ 3 eV $) to become much closer to the Fermi energy. However, under ambient conditions, the $4s$-like band is not significantly moved in CSC DFT+DMFT calculations, which makes sense because of the minimum hybridization between the $4s$-like band and the group of $p$-like and $d$-like bands (Fig. \ref{fig:NiO_fat_bands}). Given the fact that a non-CSC scheme was successfully applied in many studies of the late TM monoxides \cite{PhysRevLett.99.156404,PhysRevB.85.245110,PhysRevB.82.195101}, we carry out the calculations in the non-CSC scheme even when the $4s$-like bands of FeO and MnO enter the correlation window (Fig.\ref{fig:4s_at_Gamma_point}). Though the $4s$-like band does not take part in the DMFT iteration, it is used in constructing the final lattice Green's function for complex band analysis. 


\subsection{Evaluation of CBS and Decay Rate}

As mentioned in the Introduction, the wavefunction decay rate in direction $\vec{k}$ can be estimated by supplying an imaginary part to it, i.e. $\vec{k} = \vec{k}_{Re} + i\vec{k}_{Im}$, and studying the complex band structure (CBS). Here, $\vec{k}_{Re}$ is the decay direction, and $\vec{k}_{Im}$ yields $\beta$. CBS is always defined on the complex plane of $\vec{k}$ where $\vec{k}_{Re}$ lies on the real axis. A grid or path of real $\vec{k}$ points in the sense of traditional Brillouin zone sampling in DFT calculations is not defined here. Contrary to the normal procedure of solving for eigenenergies after specifying real $\vec{k}$ points, one needs to first specify the value of the eigenenergy, then search the complex plane of $\vec{k}$ at that eigenenergy for poles of the full Green's function to locate the $\vec{k}$(s) for that eigenenergy. In practice, it's convenient to express any $\vec{k}$ in the 1st Brillouin zone as: 
\begin{equation}
\vec{k} = \mathcal{C}_{1} \cdot \hat{k}_{\perp} + (C_{2} \cdot \hat{k}_{2} + C_{3} \cdot \hat{k}_{3}) \\ 
\equiv \vec{k}_{\perp} + \vec{k}_{\parallel}
\end{equation}
where the \textit{real} unit vector $\hat{k}_{\perp}$ is \textit{always} the decay direction, and $\mathcal{C}_{1}$ is the \textit{complex} coefficient that defines the complex plane of searching for poles, $i.e.$ $(\mathcal{C}_{1} \cdot \hat{k}_{\perp}) \equiv \vec{k}_{\perp}$. Here, the quantity ($Re[\mathcal{C}_{1}] \cdot \hat{k}_{\perp}$) is same as the $\vec{k}_{Re}$ used at the beginning of this section. And, $\hat{k}_{2}$ and $\hat{k}_{3}$ are user-defined \textit{real} unit vectors in the plane perpendicular to $\hat{k}_{\perp}$. $C_{2}$ and $C_{3}$ are both \textit{real}. $C_{2} \cdot \hat{k}_{2} + C_{3} \cdot \hat{k}_{3} \equiv \vec{k}_{\parallel}$ defines the parallel component of $\vec{k}$. 

\begin{figure}[H]
  \centering
  \includegraphics[width=1.0\columnwidth]{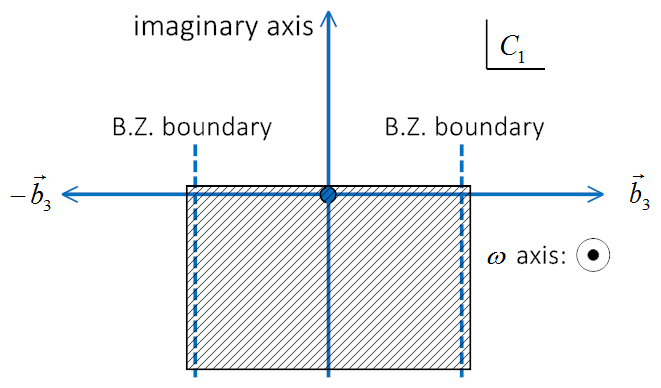}
  \caption
  {Complex plane of $\mathcal{C}_{1}$ for searching poles of $G(\mathcal{C}_{1},\omega)$, for the direction $\hat{k}_{\perp} = {\vec{b}_{3}} / {| \vec{b}_{3} |}$. Shaded area is the searching area, which includes the real axis and boundaries of the 1st Brillouin zone.} 
  \label{fig:def_area_of_CBS}
\end{figure}

The DMFT self-energy in real frequency domain is used to construct the lattice Green's function of the subspace containing $s$, $p$ and $d$ orbitals (9-dimensional). The full Green's function is expressed in the usual way:
\begin{equation}
G(\vec{k},\omega)=\frac{1}{\omega + \mu - \tilde{H}^{Wann}(\vec{k}) - \Sigma^{DMFT}(\omega)}
\end{equation}
In Eq.(6), $\mu$ is the Fermi energy, $\Sigma (\omega)$ is the converged DMFT self-energy after analytical continuation. The original $H^{Wann}(\vec{k})$ is Fourier transformed to real space hopping $t_{ij}(\vec{R})$, then Fourier transformed back to $\tilde{H}^{Wann}(\vec{k})$ at any $\vec{k}$, real or complex, as defined in Eq.(5). For given $\hat{k}_{\perp}$ and $\vec{k}_{\parallel}$, $\tilde{H}^{Wann}(\vec{k})$ can be considered as a function of $\mathcal{C}_{1}$ only. Thus the Green's function is a function of $\mathcal{C}_{1}$ and $\omega$, $i.e.$ $G(\mathcal{C}_{1},\omega)$. Fig.\ref{fig:def_area_of_CBS} shows an example case of the complex plane defined by $G(\mathcal{C}_{1},\omega)$. In the example $\hat{k}_{\perp}$ is chosen to be the unit reciprocal lattice vector ${\vec{b}_{3}} / {| \vec{b}_{3} |}$ and $\vec{k}_{\parallel}=0$. The poles of $G(\mathcal{C}_{1},\omega)$ are resolved by finding roots of the equation: $det|G^{-1}(\mathcal{C}_{1},\omega)|=0$. The set of roots for different values of $\omega$ gives the complex bands. In the following work, we study the complex band and decay rate in three directions: (a) $\hat{k}_{\perp} = (k_{x},k_{y},k_{z}) = (1,0,0)$; (b) $\hat{k}_{\perp} = (1,1,0)/\sqrt{2}$; (c) $\hat{k}_{\perp} = (1,1,1)/\sqrt{3}$, where $k_{x},k_{y},k_{z}$ are Cartesian coordinates in $k$-space. 

\section{\label{sec:3} Results and Discussion}

The main purpose is to study the features of the CBS when the correlation effect is included. In this section, we first report the band gaps and spectral functions from DFT+DMFT and compare it to experimental data and existing calculations. Then the CBS and $\beta$ parameters are studied at the so-called charge neutrality level within the band gap, which is further related to the Fermi level pining position and the height of Schottky barrier in tunnel junction applications. 

\subsection{ Band Gaps from DFT+DMFT } 

\begin{figure}[H] 
  
  \begin{subfigure}[b]{0.5\linewidth} 
    \centering
    \includegraphics[width=1.0\columnwidth]{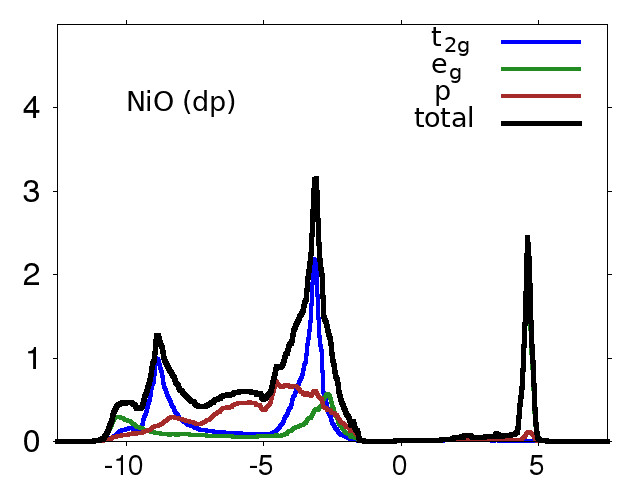} 
  \end{subfigure}
  \begin{subfigure}[b]{0.5\linewidth}
    \centering
    \includegraphics[width=1.0\columnwidth]{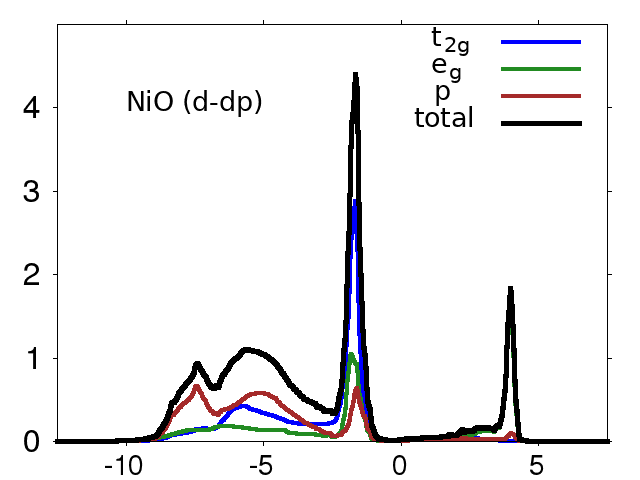} 
  \end{subfigure} 
  
  \begin{subfigure}[b]{0.5\linewidth}
    \centering
    \includegraphics[width=1.0\columnwidth]{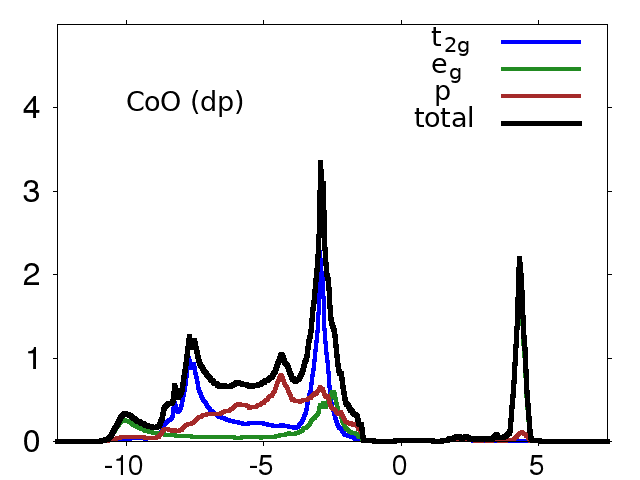} 
  \end{subfigure}
  \begin{subfigure}[b]{0.5\linewidth}
    \centering
    \includegraphics[width=1.0\columnwidth]{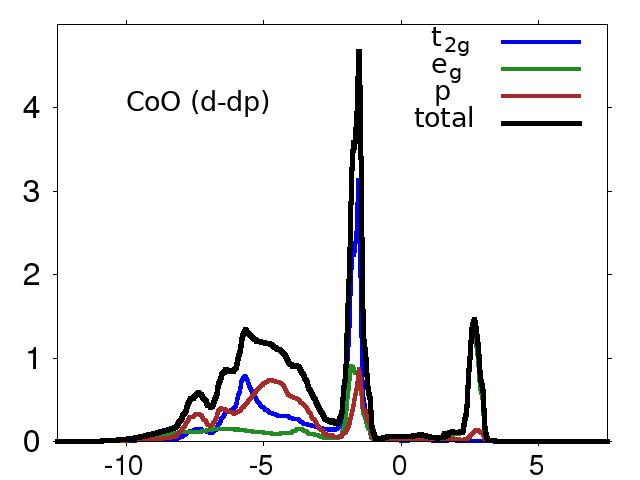} 
  \end{subfigure}
  
  \begin{subfigure}[b]{0.5\linewidth}
    \centering
    \includegraphics[width=1.0\columnwidth]{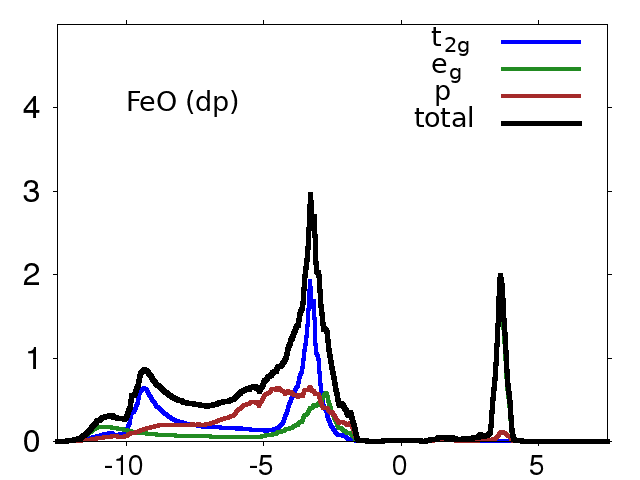} 
  \end{subfigure}
  \begin{subfigure}[b]{0.5\linewidth}
    \centering
    \includegraphics[width=1.0\columnwidth]{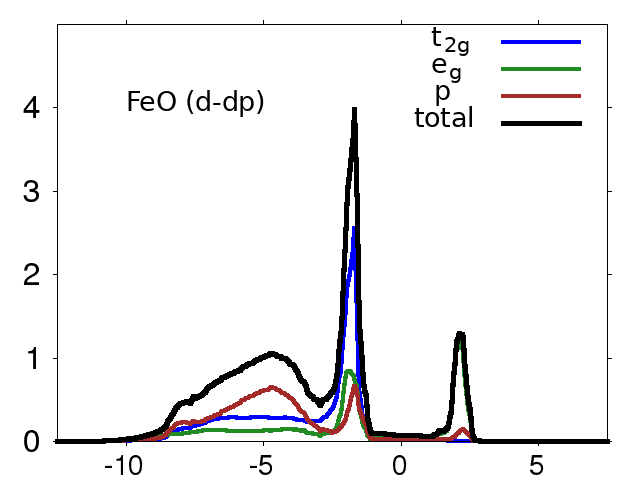} 
  \end{subfigure}
  
  \begin{subfigure}[b]{0.5\linewidth}
    \centering
    \includegraphics[width=1.0\columnwidth]{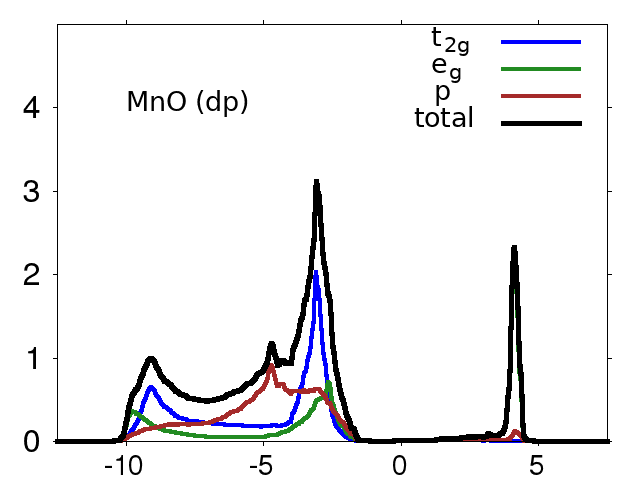} 
  \end{subfigure}
  \begin{subfigure}[b]{0.5\linewidth}
    \centering
    \includegraphics[width=1.0\columnwidth]{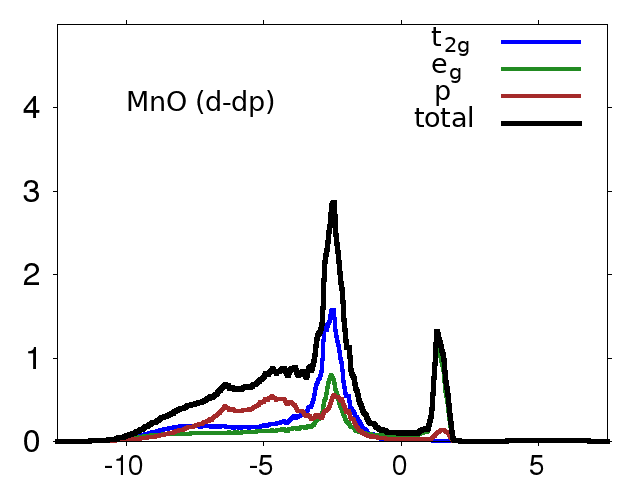} 
  \end{subfigure}
  
  \caption{Total(black), $t_{2g}$(blue), $e_{g}$(green) and $p$(red) spectral functions, $A(\omega)$, of the two models of NiO, CoO, FeO and MnO.}
  \label{fig:spectral_dp_and_ddp} 
\end{figure}

The resulting density of states (DOS) from the DMFT calculations are presented in Fig.\ref{fig:spectral_dp_and_ddp} for the $dp$ and $d$-$dp$ models. The band gaps are measured between the widths at half height of the conduction band and valence band in DOS. Table \ref{table:1} has the measured band gaps. Here we briefly compare to the experiments. NiO and MnO have been extensively studied in experiments. The band gap of NiO was determined to be $3.7$--$4.5 \eV$ \cite{band_gap_exp_NiO_CoO, band_gap_exp_NiO}. The experimental values of MnO gap is in the range of $3.6$--$4.0 \eV$ \cite{band_gap_exp_CoO_MnO,band_gap_exp_NiO_CoO}. The present calculation of the $d$-$dp$ model are in agreement with experiments for these two materials. Experimental values of the band gap of CoO has diverse values. Some experiments reported a band gap of $2.5$--$2.8 \eV $ \cite{band_structure_sX-LDA_NiOCoOFeOMnO}. Some other studies found higher values, e.g. $5.4 \eV $ gap based on ellipsometry spectra data \cite{optical_response_in_far-infrared_regime_NiOCoOFeOMnO} and indirect gap of $ 2.8 \eV $ and direct gap of $ 5 \eV $ based on absorption spectrum from measured dielectric function \cite{band_structure_sX-LDA_NiOCoOFeOMnO}. The value of our calculated band gap based on the $d$-$dp$ model falls in the range and is close to the conductivity experiment \cite{GVISHI1972893}. Quasiparticle calculations using the hybridfunctional and the $G_{0}W_{0}$ method yields similar value ($ 3.4 \eV $) for the band gap \cite{PhysRevB.79.235114} of CoO. It seems the absorption experiments underestimates the gap compared to DMFT results. Unfortunately, there are very limited experiments reporting the measured band gaps of FeO in PM phase. This is because the preparation of a pure FeO sample is difficult due to Fe segregation \cite{band_gap_exp_FeO}. Thus, comparison of theoretical calculation with experimental spectra is very rare. The only reported experimental estimate that we found is $ 2.4 \eV $ from an optical absorption measurement \cite{band_gap_exp_FeO}. We are not sure if this is an underestimated value. 

    \begin{table}[H]
    \fontsize{8}{14}\selectfont
    \centering
    \begin{tabular}{ | >{\centering}m{4cm} | c | c | c | c |}
    \hline
    Band Gap (eV)                & NiO  & CoO  & FeO  & MnO   \\ \hline 
    DFT+DMFT,     $dp$ $model$   & 6.74 & 6.63 & 6.05 & 6.74  \\ \hline 
    DFT+DMFT, $d$-$dp$ $model$   & 4.76 & 3.72 & 3.37 & 3.72  \\ \hline
    Exp(conductivity)            & 3.7  & 3.6  & n/a  & 3.8   \\ 
    Exp(XAS-XES)                 & 4.0  & 2.6  & n/a  & 4.1   \\ 
    Exp(PES-BIS)                 & 4.5  & 2.5  & n/a  & 3.9   \\     
    Exp(absorption)              & 4.0  & 2.8  & 2.4  & 3.6-3.8  \\ \hline   
    \end{tabular}
    \caption{The band gaps measured from DFT+DMFT density of states. Sources of the experimental gaps are in text of Sec.III.A.}
    \label{table:1}
    \end{table}
    
%

We noticed our band gap result of the $d$-$dp$ model of NiO is in general agreement with existing DFT+DMFT calculations. Though the existing calculations are not exactly same, most of them are performed in the PM state and are not in CSC scheme. In Ref.\cite{Korotin2008} a similar model construction was considered and $U=6.6 \eV$ (no $J$ involved) calculated from constrained-LDA was used. A band gap of about $4 \eV$ was obtained. In Ref.\cite{PhysRevB.74.195114}, only the Ni-$d$ orbitals were taken into account (which should be called $d$-$d$ model following the naming convention used here) and $U= 8.0 \eV $ and $J=1.0 \eV $ were used in the calculation. They calculated a band gap of $4.3 \eV$ for NiO. Ref.\cite{M.Karolak_double_counting_NiO} had used NiO to test a new double counting method, where they used the same $U= 8.0 \eV$ and $J = 1.0 \eV$ and involved both $d$ and $p$ orbitals. The calculation was carried at high T of ~2300K. They found a band gap of about $ 4.3 \eV $ as well. Within this study, both the size of band gap and the position of the $p$ orbital peak below Fermi energy depend on the double counting potential. 

\begin{figure}[H] 
  \begin{subfigure}[b]{0.5\linewidth}
    \centering
    \includegraphics[width=1.0\columnwidth]{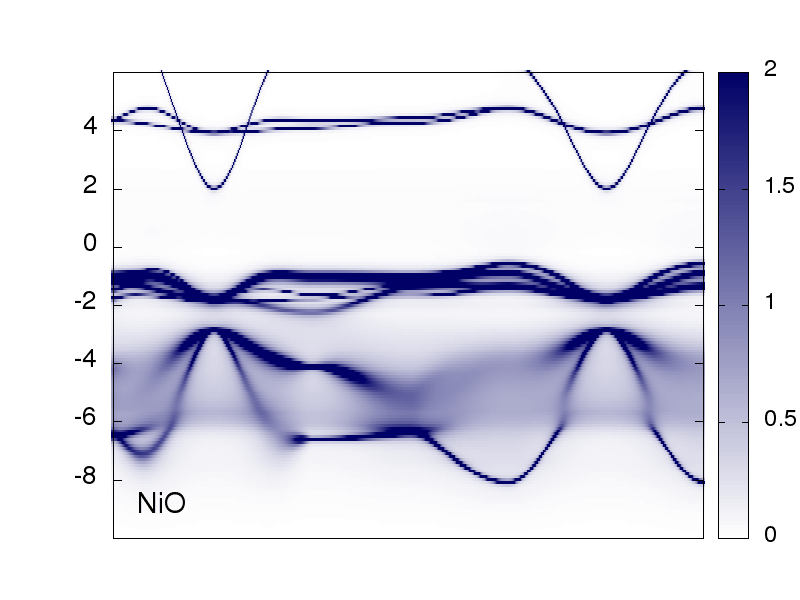}
    \caption{NiO} 
  \end{subfigure}
  \begin{subfigure}[b]{0.5\linewidth}
    \centering
    \includegraphics[width=1.0\columnwidth]{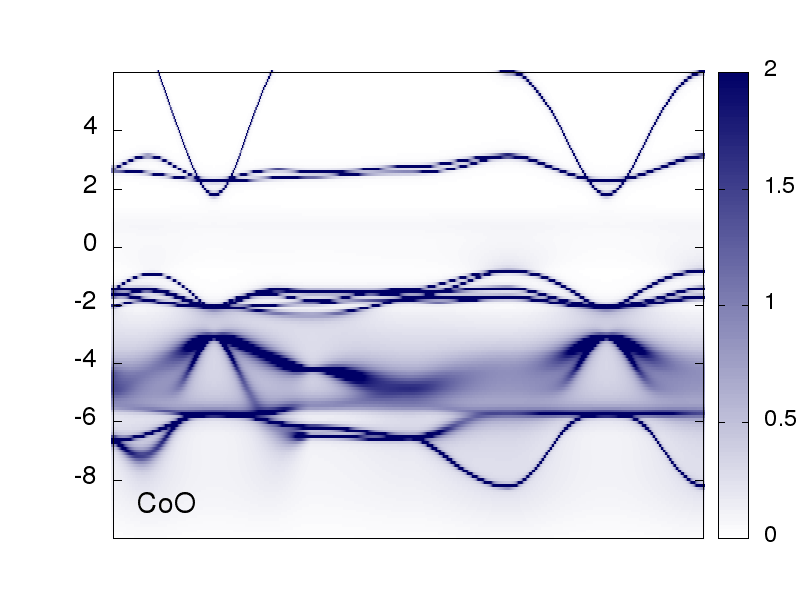}
    \caption{CoO} 
  \end{subfigure} 
  \begin{subfigure}[b]{0.5\linewidth}
    \centering
    \includegraphics[width=1.0\columnwidth]{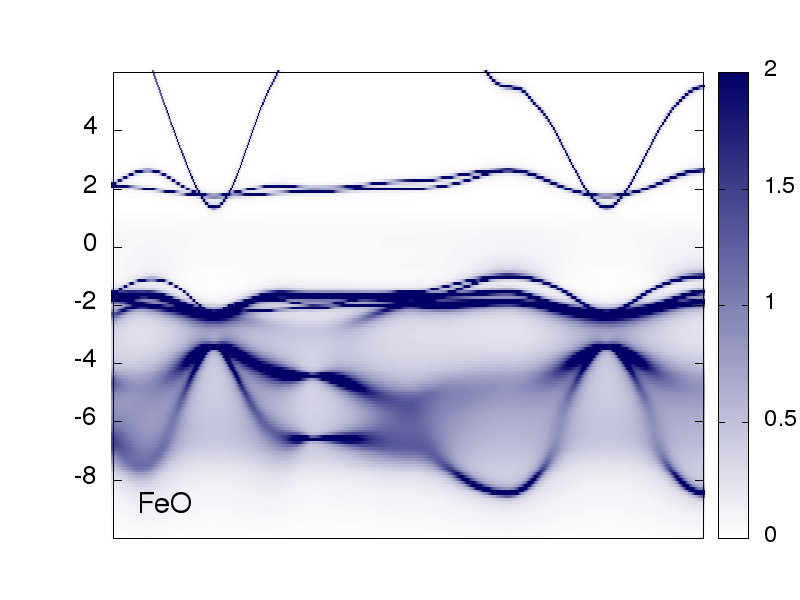}
    \caption{FeO} 
  \end{subfigure}
  \begin{subfigure}[b]{0.5\linewidth}
    \centering
    \includegraphics[width=1.0\columnwidth]{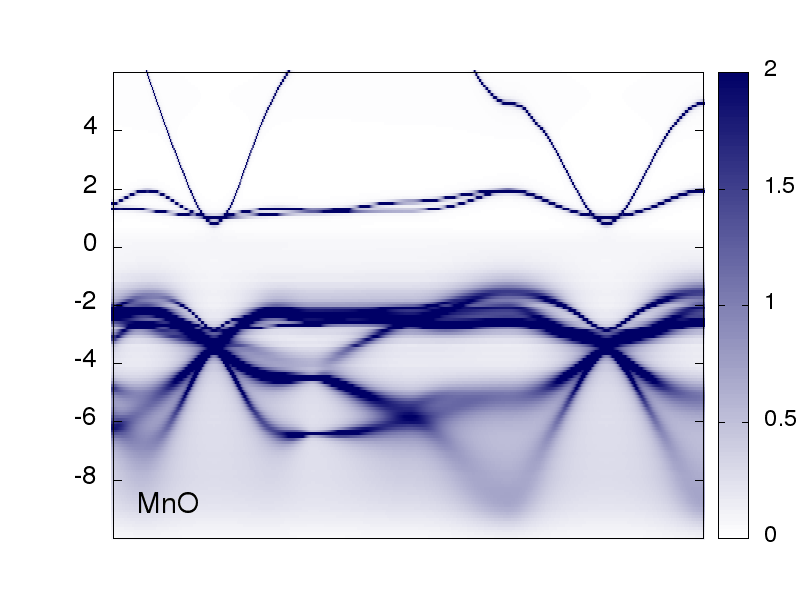}
    \caption{MnO} 
  \end{subfigure}
  \caption{$\vec{k}$-resolved spectral functions of the $d$-$dp$ model, of NiO, CoO, FeO and MnO.}
  \label{fig:k_resolved_spectral_dp_and_ddp} 
\end{figure}

In our calculation, the correct representation of the $U$ matrices is important to yield accurate band gaps for the four materials. We have found results from the $d$-$dp$ model are in better agreement with existing experiments than the $dp$ model, which seems to overestimate the band gap. However, the preference of the $d$-$dp$ model over the $dp$ model should be more carefully supported by taking into consideration of other important factors. For example, it has been proposed that the current cRPA method could be improved by including the Pauli exclusion principle in the formalism, and overall the effect would be a reduction of the interaction strength \cite{PRB_accuracy_of_crpa}. There are also other factors within the DMFT, e.g. different double counting methods, that affect the resulting band gap and spectral function. For example the double counting methods introduced in Ref.\cite{M.Karolak_double_counting_NiO} and Ref.\cite{20.500.11850/155421}. Those factors are worthy of dedicated studies and are outside the scope of the current work. Thus we limit our discussion to be within the original $d$-$dp$ cRPA scheme and with only fully localized limit (FLL) double counting in the DMFT part. The $dp$ cRPA scheme is not used in later decay rate analysis, but exists in the Appendix for purpose of benchmarking the cRPA implementation. We would like to emphasize that, for consistency, one set of projected Wannier orbital basis functions is used in both the downfolded Hamiltonian $H(k)$ and the cRPA-calculated Coulomb $U$ matrices, which is different from existing calculations of these materials. 

In order to include the effect of the TM $4s$ band in the later complex band analysis, we prepare the full Green's function, Eq.~(6), in the Wannier orbital basis containing $d$, $p$, and $s$ orbitals. The DMFT self energy is associated with the $d$ orbitals. The $k$-resolved spectral function $\mathcal{A}(\vec{k},\omega) = (-1/\pi) \, \Im[G(\vec{k},\omega)]$ of the $d$-$dp$ model is shown in Fig.~\ref{fig:k_resolved_spectral_dp_and_ddp}. The rest of the calculations are based on this Green's function. 

\subsection{ Complex band structure including DMFT self-energy } 

The complex band structure (CBS) and real band structure (RBS) are obtained from resolving poles of the Green's function, as explained in Section II. In this section we analyze the obtained complex band structure and argue about the pinning position of the Fermi level for tunnel junction applications and calculating the $\beta$ parameter at that energy level.  

In a general tunnel junction setup, where the insulating material is connected to metal leads on both sides, the Fermi levels of the two materials are brought into coincidence. At the interface, there are metal-induced gap states (MIGS) \cite{PhysRev.138.A1689} in the insulator gap region decaying exponentially into the material, which are Bloch states of the insulator with complex wave vector. For semiconductors, this forms a continuum of states around the Fermi energy ($E_{F}$) within the gap. The gap states continuously change (as function of energy) from valence- to conduction-band character, appearing as an arch-shaped complex band connecting the real valence band with the real conduction band. The idea of local \textit{charge neutrality}, proposed by Tersoff \cite{PhysRevLett.52.465}, says: when the density of MIGS is reasonably large, it is necessary to occupy those MIGS which are primarily valence-band character, while leaving those of mainly conduction-band character empty. Therefore, $E_{F}$ should be pinned at or near the energy where the gap states cross over from valence- to conduction-band character. If the complex band within the gap is a smooth curve, the crossing over is naturally found at $dE/dk \rightarrow \infty$, which is called the \textit{branch point}, and the corresponding energy is called the charge neutrality level $E_{B}$ \cite{0022-3719-10-12-022,PhysRevLett.52.465,FLORES1992301}. 

\begin{figure}[H] 
  \centering
  \begin{subfigure}[b]{0.8\linewidth}
    \centering
    \includegraphics[width=1.0\columnwidth]{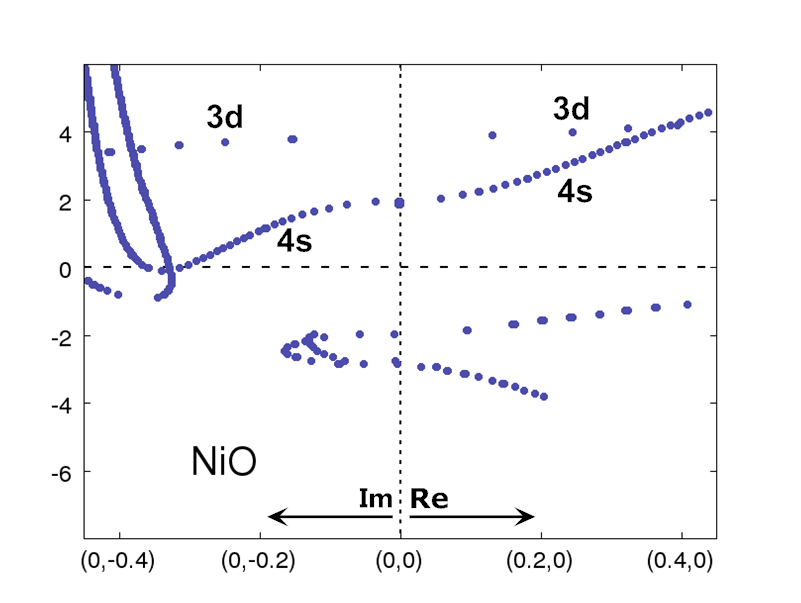}
  \end{subfigure}
  
  \begin{subfigure}[b]{0.8\linewidth}
    \centering
    \includegraphics[width=1.0\columnwidth]{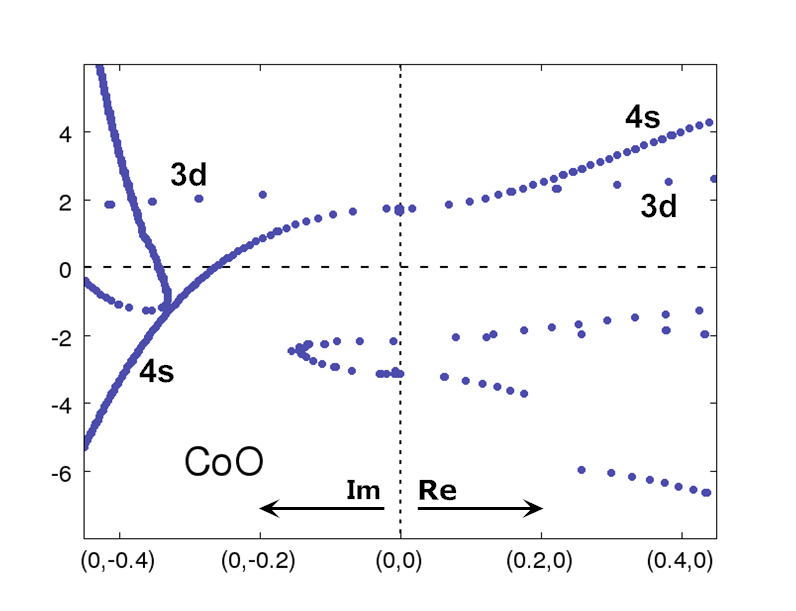}
  \end{subfigure} 
  
  \begin{subfigure}[b]{0.8\linewidth}
    \centering
    \includegraphics[width=1.0\columnwidth]{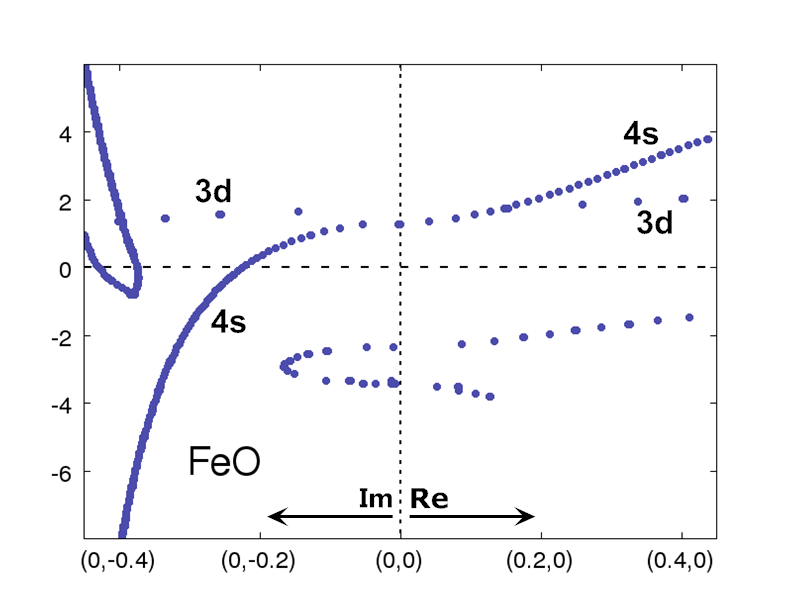}
  \end{subfigure}

  \begin{subfigure}[b]{0.8\linewidth}
    \centering
    \includegraphics[width=1.0\columnwidth]{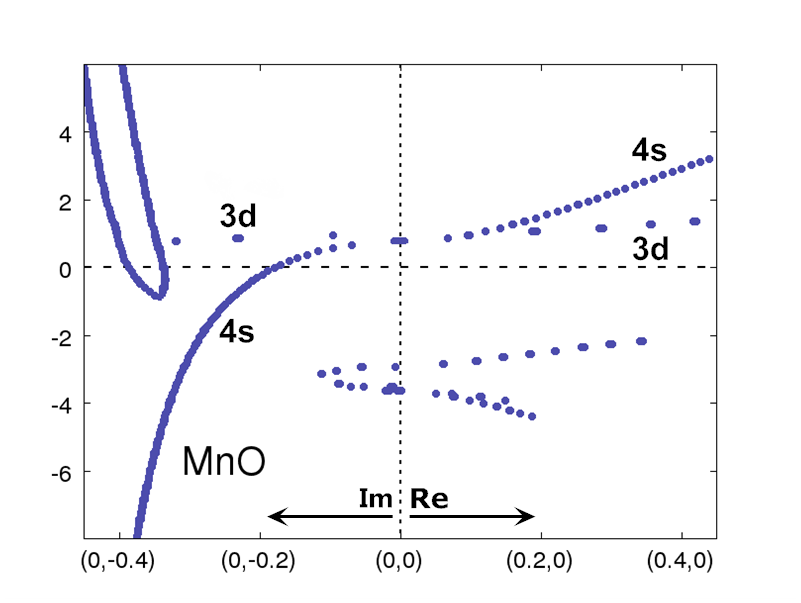}
  \end{subfigure}
  
  \caption{Complex band structures of the $d$-$dp$ model. The decay direction being studied is $\hat{k}_{\perp} = {\vec{b}_{1}} / {| \vec{b}_{1} |}$, with $C_{2}=C_{3}=0$ (see Eq.(5) for definition). The path for the band plot is: $(Re[\mathcal{C}_{1}],Im[\mathcal{C}_{1}])=(0,-0.5) \rightarrow (0,0) \rightarrow (0.5,0)$. The DFT+DMFT Fermi level, $E_{F}^{DMFT}$, is at zero. }
  \label{fig:CBS_and_RBS}
  
\end{figure}
\begin{figure}[H]
  \centering
  \includegraphics[width=1.0\columnwidth]{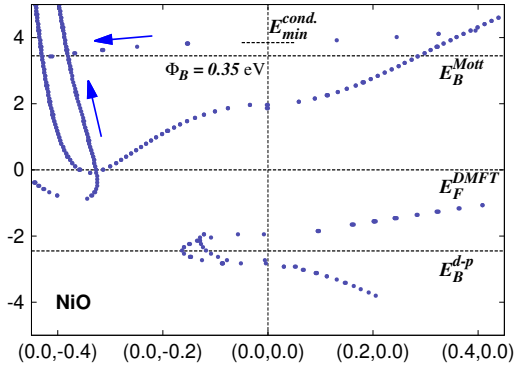}
  \caption{The CBS of the $d$-$dp$ model of NiO. $x$-axis is the path of $(Re[\mathcal{C}_{1}],Im[\mathcal{C}_{1}])$. $E_{F}^{DMFT}$ is at zero (eV). $E_{B}^{\Mott}$ is located at the crossing of the imaginary extension of conduction band (small slope arrow) and the imaginary extension of valence band (large slope arrow). $E_{F}^{DMFT}$ is at zero. $E_{B}^{d-p}$ is found within the imaginary band connecting $d$- and $p$-valence bands where $dE/dk \rightarrow \infty$. The conduction band minimum $E_{\min}^{\cond}$ (short horizontal dash line) is found at $\Gamma$ at a slightly higher energy than $E_{B}^{\Mott}$. The Schottky barrier height is measured to be: $\Phi_{B} = E_{\min}^{\cond} - E_{B}^{\Mott} \approx 0.35 \eV.$}
  \label{fig:branch_points}
\end{figure}

Fig.\ref{fig:CBS_and_RBS} displays the CBS and RBS for $\hat{k}_{\perp} = {\vec{b}_{1}} / {| \vec{b}_{1} |}$ and $\vec{k}_{\parallel}=0$, where $\vec{b_{1}}$ is a reciprocal lattice vector. The CBS part, $(Re[\mathcal{C}_{1}],Im[\mathcal{C}_{1}]) = (0.0,-0.5) \rightarrow (0.0,0.0)$ in the left half of Fig.\ref{fig:CBS_and_RBS}, contains important information for determining the charge neutrality level $E_{B}$ and wavefunction decay rate $\beta$. One major observation is that the gap states transitioning from below to above the Mott gap are not continuous. The real conduction bands and the top of valence bands are both primarily $d$-orbitals, as seen in the DOS. The extensions of these bands into the complex sector cross each other rather than connect smoothly, which is different from typical semiconductors. For NiO in Fig.\ref{fig:CBS_and_RBS}, the crossing happens at about $ 3.5 \eV $ above the DFT+DMFT Fermi level ($E_{F}^{DMFT}$) and $\mathcal{C}_{1} = (0.00,-0.37)$. Assuming the charge neutrality condition still applies, the gap states of valence-band character are occupied and $E_{B}$ should be pinned at or near the crossing point, where $dE/dk$ has a finite jump rather than $\rightarrow \infty$. This is more clearly shown in Fig.\ref{fig:branch_points}. We call this crossing point the Mott branch point, and the corresponding energy level $E_{B}^{\Mott}$. The conduction band obtained from DFT+DMFT is very flat. Its extension in the complex domain is almost a straight line of small slope, while the complex band originated from valence bands goes up steeply from below the gap. This feature leads to the $E_{B}^{\Mott}$ being very close to the conduction band minimum $E_{\min}^{\cond}$ (which is found at $\Gamma$ in our case). The small difference between $E_{B}^{\Mott}$ and $E_{\min}^{\cond}$ gives a small Schottky barrier height $\Phi_{B}$ of the tunnel junction, i.e. $\Phi_{B} = E_{\min}^{\cond} - E_{B}^{\Mott}$. Our calculation implies NiO is a junction material with large gap and small barrier height (comparing with the ~$1.0$--$1.5 \eV $ band gap and ~$0.7$--$1.0 \eV $ barrier height of typical semiconductors like Si and GaAs). We found $\Phi_{B}^{NiO} \approx 0.35 \eV $ from the calculated CBS of the $d$-$dp$ model of NiO, as indicated in Fig.\ref{fig:branch_points}. There is not much reported work on tunnel junction experiments using late transition metal monoxides. We found the most relevant work was done on the Ni-NiO-Ni tunnel junction \cite{Hobbs_Ni_NiO_Ni_junction}, where the system was found to be a very-low-barrier system with $\Phi_{B} \approx 0.2 \eV $. The barrier height varies slightly from $ 0.22 \eV $ at 4 K to $ 0.19 \eV $ at 295 K. This is in qualitative agreement with our calculation. Ni was used as metal leads for easy manufacture. It is possible to tune the composition of the metal and oxide so as to raise the barrier height slightly. 

The $4s$ band plays an interesting role here. By following the path: $(Re[\mathcal{C}_{1}],Im[\mathcal{C}_{1}])=(0.5,0.0) \rightarrow (0.0,0.0) \rightarrow (0.0,-0.5)$, we observe that the real $4s$ band goes from high to low energy at the $\Gamma$ point, which is a common feature in Fig.\ref{fig:CBS_and_RBS}. NiO presents a different feature in the complex sector: it first decreases to about the Fermi energy and then starts to increase to higher energy, while the $4s$ bands of the other three materials decrease in energy monotonically. In the case of NiO in Fig.\ref{fig:branch_points}, the complex $4s$ band crosses the complex extension of valence $d$ band at a point very close to $E^{DMFT}_{F}$, and the real 4$s$ band has $ E_{\min}^{\cond} \approx 2 \eV $ at $\Gamma$. By using the $4s$ band instead of the conduction $3d$ band, one keeps the Fermi level pinned very close to $E^{DMFT}_{F}$ (\textit{i.e.} there would be nearly no shift in a tunnel junction), and $\Phi_{B} \approx 2 \eV $ for NiO, which is too far away from experiment. We believe the $4s$ band is not responsible for correctly determining the Schottky barrier height also because the orbital character should not suddenly change (from $3d$ to $4s$) at the branch point. The correct $E_{B}^{\Mott}$ in Fig.\ref{fig:branch_points} is located by following the same $d$ orbital character. The same argument applies to CoO, where the complex $4s$ band is just touching the complex valence $d$ band at a point close to the gap bottom. CoO should also be a very small barrier junction material. The $4s$ bands of FeO and MnO fall into an area where no other complex band is found. We can expect them to be small-barrier materials based on \textit{only} complex $d$ bands.

It is worthwhile to mention that we do observe a continuous transition of "gap" states below $E_{F}^{DMFT}$, where real $p$-orbital bands and $d$-orbital bands are connected by continuous complex bands. This is also shown in Fig.\ref{fig:branch_points}. The point where $dE/dk \rightarrow \infty$ can be located, and we call it the $d$-$p$ branch point. The corresponding energy level is $E_{B}^{d-p}$. The feature actually exists in the DFT calculation alone, despite the fact that it yields a metallic ground state. The feature largely remains after the DMFT calculation. The energy range of the arch-shaped complex band containing $E_{B}^{d-p}$ may decrease due to the gap opened by DMFT, as seen in the CBS of CoO, FeO and MnO in Fig.\ref{fig:CBS_and_RBS}. 

Numerically some regions of the complex and real bands smear out, for example near $\Gamma$ in energy range of $[-7,-3] \eV$ below $E_{F}^{DMFT}$. Such smearing is explained within the quasi-particle picture in which spectral weights have spreads, as seen in the $\vec{k}$-resolved spectral functions in Fig.\ref{fig:k_resolved_spectral_dp_and_ddp}. 


\vspace{2mm}
In the rest of this section, we turn to the calculation of $\beta$ at $E_{B}^{\Mott}$ for the $d$-$dp$ model of the four materials and study the direction dependence of $\beta$. 
\vspace{2mm}


The values of the decay rate $\beta$ should be anisotropic for crystal structures. As already mentioned in the end of Sec.II.C, we perform the calculation of $\beta$ in three decay directions: (a) $\hat{k}_{\perp}=(1,0,0)$, (b) $\hat{k}_{\perp} = (1,1,0)/ \sqrt{2}$ and (c) $\hat{k}_{\perp} = (1,1,1)/ \sqrt{3}$. The values $E_{B}^{\Mott}$ and $E_{B}^{d-p}$ are determined for each direction. We found $E_{B}^{\Mott}$, or $E_{B}^{d-p}$, stays the same for the three different decay directions. If the wavefunction propagates perfectly along one direction, then we get the decay rate by directly reading off $Im[\mathcal{C}_{1}]$ (thus $\beta \equiv | 2 \cdot Im[\mathcal{C}_{1}] | $) at the branch point. This has been done for the three directions and for the four materials. The results are summarized in Table \ref{table:2}. We found the values of $\beta$ at $E_{B}^{\Mott}$ are all within the range of [0.29,0.40].

    \begin{table}[H]
    \fontsize{8}{14}\selectfont
    \centering
    \begin{tabular}{ | >{\centering}m{1.5cm} | >{\centering}m{1.5cm} | c | c | c | c | }
    \hline 
    \multicolumn{2}{|c|}{ }  & NiO  & CoO  & FeO  & MnO \\  
    \hline \hline   
    \multicolumn{2}{|c|}{$\Phi_{B}=E_{\min}^{\cond}-E_{B}^{\Mott}$ (eV)} & 0.35 & 0.32 & 0.27 & 0.30 \\ 
    \hline \hline   
    \multicolumn{2}{|c|}{$E_{B}^{\Mott}$ (eV, from $E_{F}^{DMFT}$)} & +3.5 & +2.0 & +1.5 & +1.1 \\ 
    \hline \hline
    \multirow{3}{*}{$\beta @ E_{B}^{\Mott}$} 
    &  $\hat{k}_{\perp}$ : (0,0,1)  &  0.29  &  0.31  &  0.32  &  0.29 \\
    &  $\hat{k}_{\perp}$ : (0,1,1)  &  0.33  &  0.34  &  0.35  &  0.33 \\
    &  $\hat{k}_{\perp}$ : (1,1,1)  &  0.37  &  0.37  &  0.40  &  0.38 \\
    \hline \hline
    \multicolumn{2}{|c|}{$E_{B}^{d-p}$ (eV, from $E_{F}^{DMFT}$)} & -2.5 & -2.3 & -2.8 & -3.0 \\ 
    \hline 
    \multirow{3}{*}{$\beta @ E_{B}^{d-p}$} 
    &  $\hat{k}_{\perp}$ : (0,0,1)  &  0.15  &  0.14  &  0.16  &  0.13 \\
    &  $\hat{k}_{\perp}$ : (0,1,1)  &  0.17  &  0.16  &  0.16  &  0.13 \\
    &  $\hat{k}_{\perp}$ : (1,1,1)  &  0.18  &  0.16  &  0.17  &  0.14 \\
    \hline
    \end{tabular}
    \caption{The calculated values of $\Phi_{B}$ and locations of $E_{B}^{\Mott}$ and $E_{B}^{d-p}$, and the decay rates $\beta$ in different directions at the corresponding energy levels.}
    \label{table:2}
    \end{table}

\begin{figure}[H] 

  \begin{subfigure}[b]{0.5\linewidth}
    \centering
    \includegraphics[width=1.0\columnwidth]{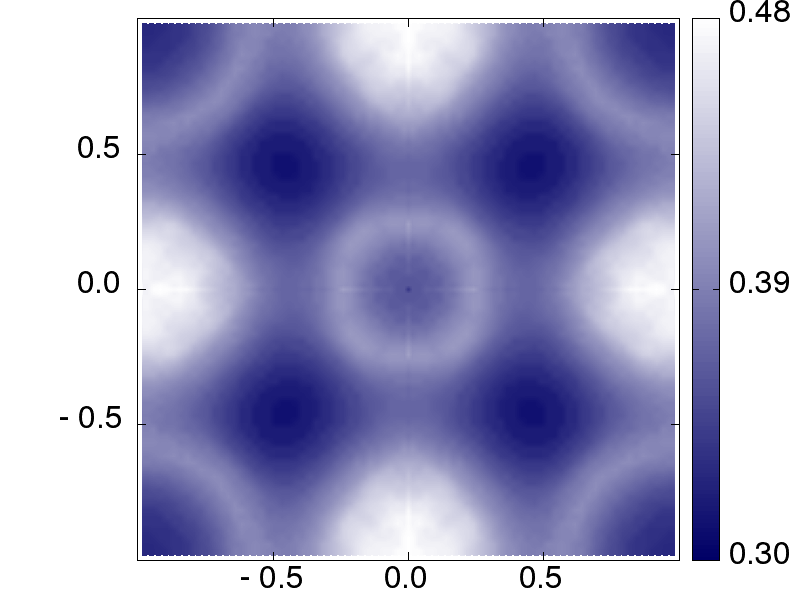} 
    \caption{NiO} 
  \end{subfigure}
  \begin{subfigure}[b]{0.5\linewidth}
    \centering
    \includegraphics[width=1.0\columnwidth]{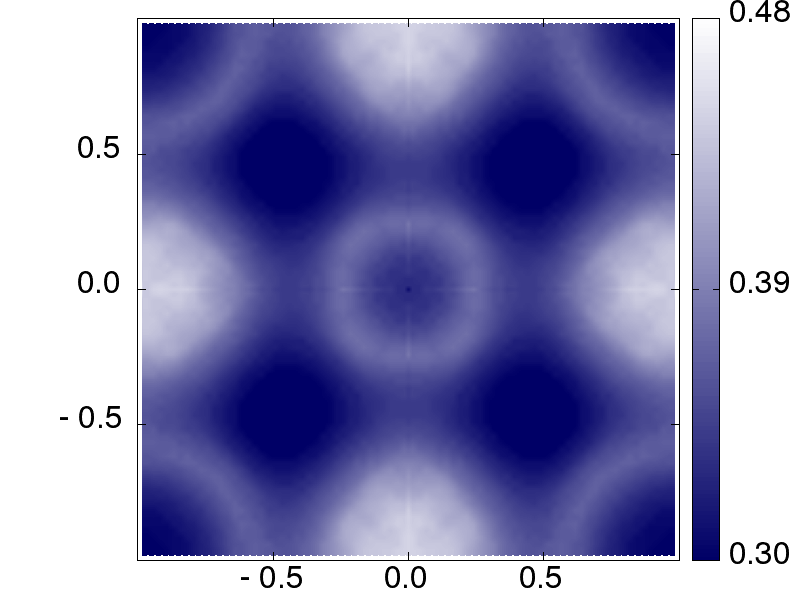} 
    \caption{CoO} 
  \end{subfigure} 
  
  \begin{subfigure}[b]{0.5\linewidth}
    \centering
    \includegraphics[width=1.0\columnwidth]{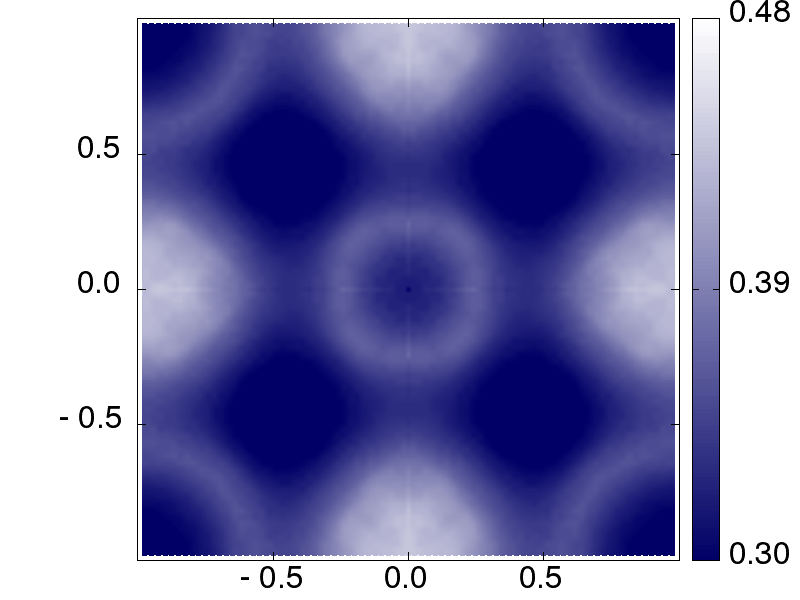} 
    \caption{FeO} 
  \end{subfigure}
  \begin{subfigure}[b]{0.5\linewidth}
    \centering
    \includegraphics[width=1.0\columnwidth]{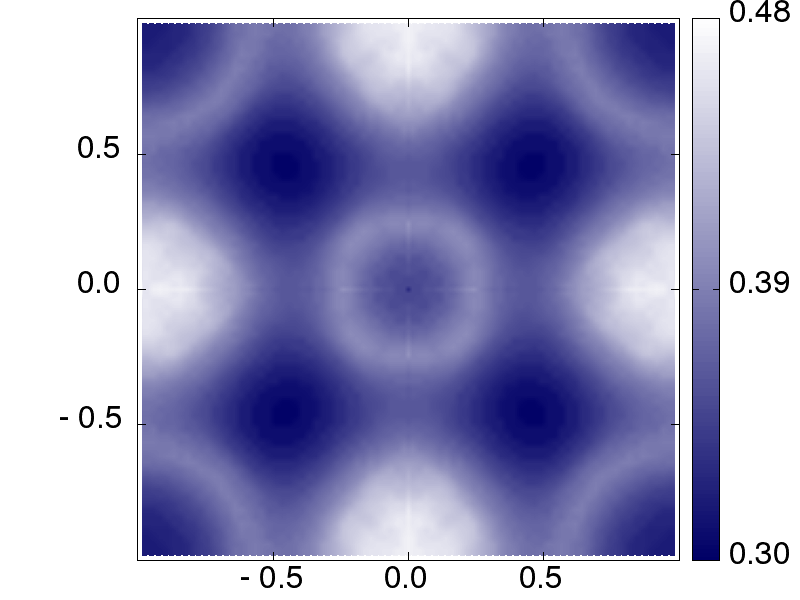} 
    \caption{MnO} 
  \end{subfigure}
  
  \caption{The distribution of $\beta(C_{2},C_{3})$ at $E_{B}^{\Mott}$ for the decay direction (1,0,0).}
  \label{fig:beta_distribution_100} 
\end{figure}

\begin{figure}[H] 

  \begin{subfigure}[b]{0.5\linewidth}
    \centering
    \includegraphics[width=1.0\columnwidth]{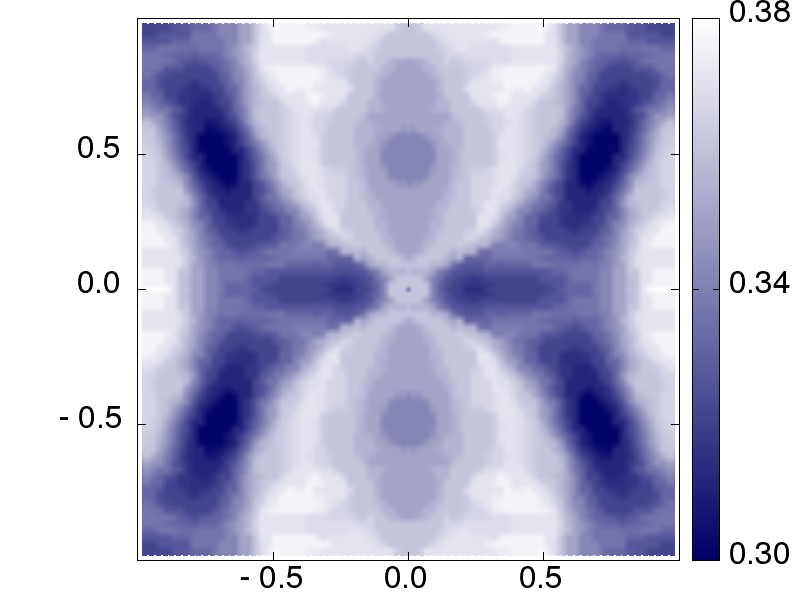} 
    \caption{NiO} 
  \end{subfigure}
  \begin{subfigure}[b]{0.5\linewidth}
    \centering
    \includegraphics[width=1.0\columnwidth]{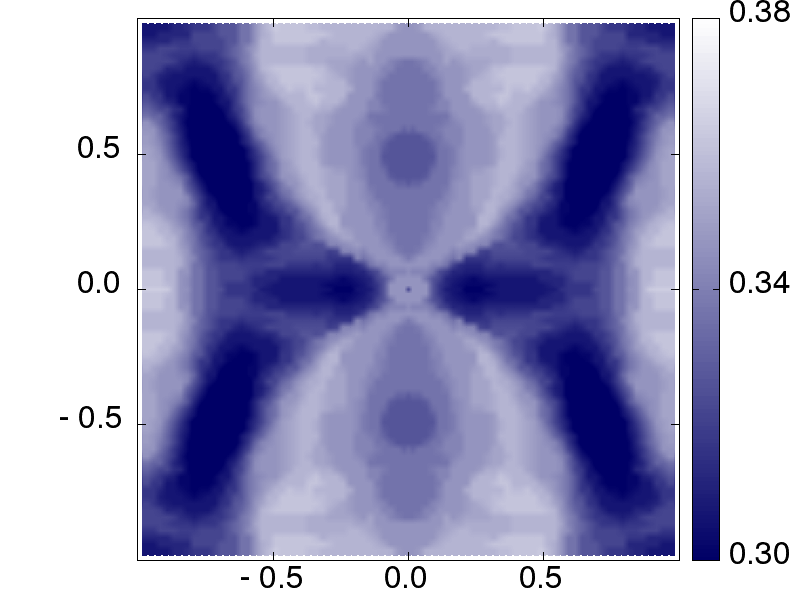} 
    \caption{CoO} 
  \end{subfigure} 

  \begin{subfigure}[b]{0.5\linewidth}
    \centering
    \includegraphics[width=1.0\columnwidth]{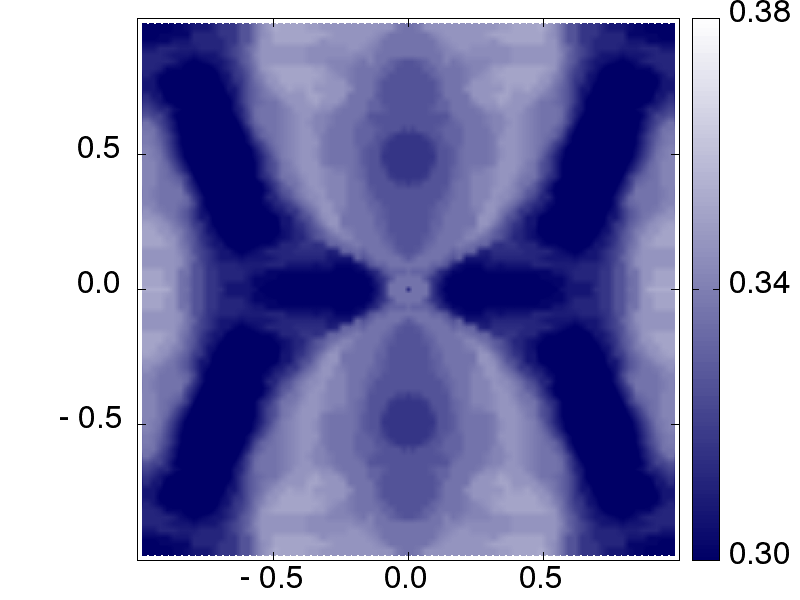} 
    \caption{FeO} 
  \end{subfigure}
  \begin{subfigure}[b]{0.5\linewidth}
    \centering
    \includegraphics[width=1.0\columnwidth]{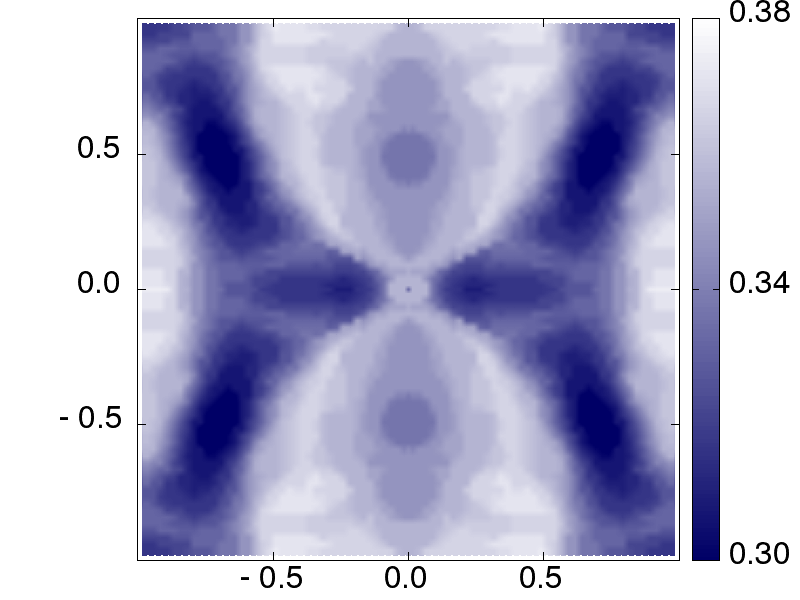} 
    \caption{MnO} 
  \end{subfigure}
  
  \caption{The distribution of $\beta(C_{2},C_{3})$ at $E_{B}^{\Mott}$ for the decay direction (1,1,0).}
  \label{fig:beta_distribution_110} 
\end{figure}

\begin{figure}[H] 

  \begin{subfigure}[b]{0.5\linewidth}
    \centering
    \includegraphics[width=1.0\columnwidth]{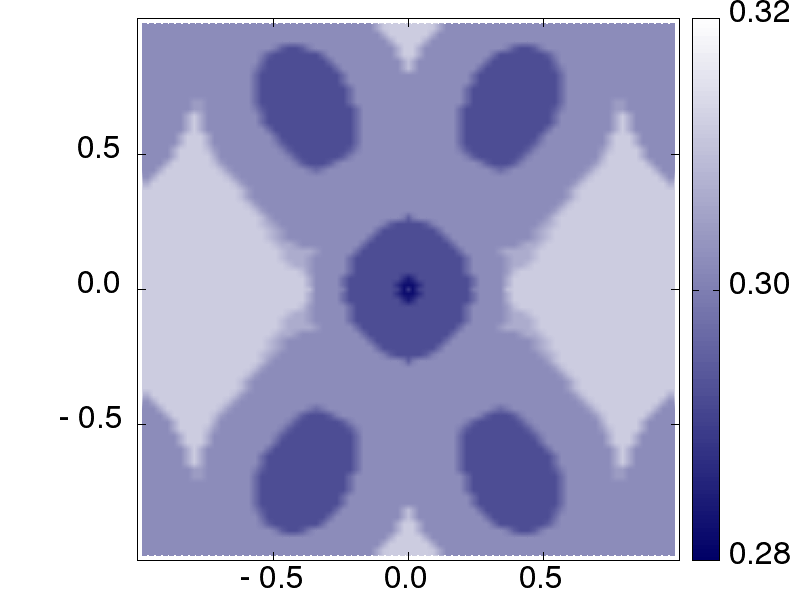} 
    \caption{NiO} 
  \end{subfigure}
  \begin{subfigure}[b]{0.5\linewidth}
    \centering
    \includegraphics[width=1.0\columnwidth]{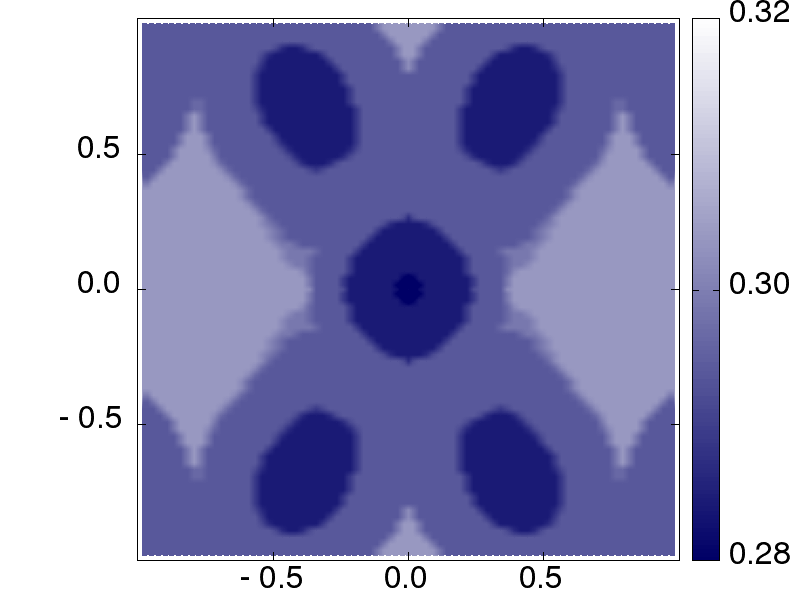} 
    \caption{CoO} 
  \end{subfigure} 
  
  \begin{subfigure}[b]{0.5\linewidth}
    \centering
    \includegraphics[width=1.0\columnwidth]{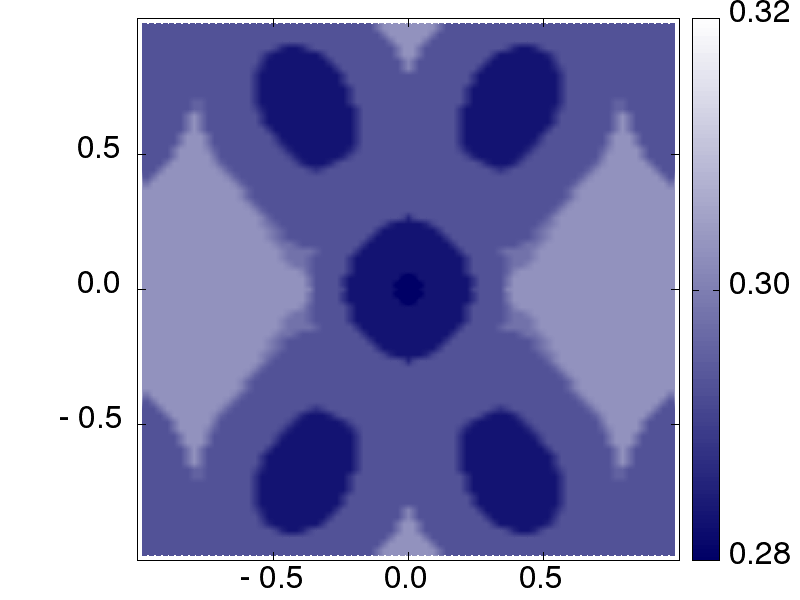} 
    \caption{FeO} 
  \end{subfigure}
  \begin{subfigure}[b]{0.5\linewidth}
    \centering
    \includegraphics[width=1.0\columnwidth]{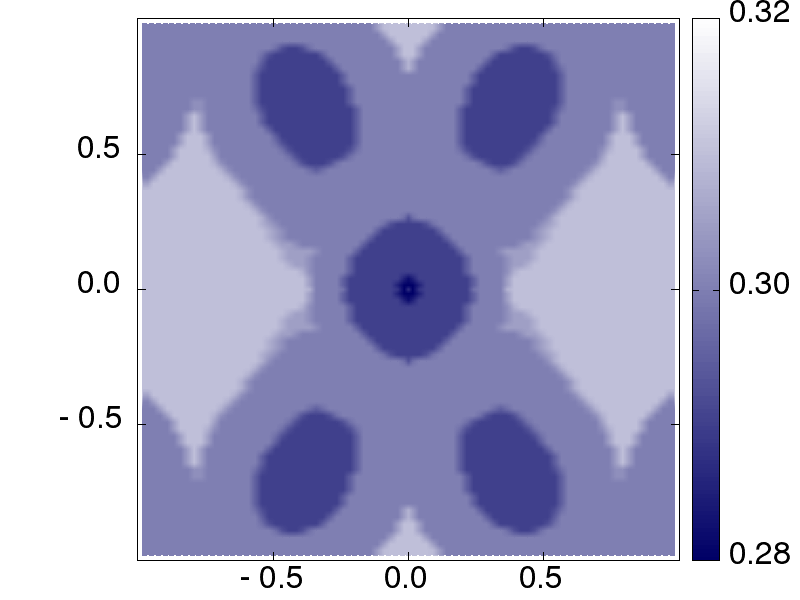} 
    \caption{MnO} 
  \end{subfigure}
  
  \caption{The distribution of $\beta(C_{2},C_{3})$ at $E_{B}^{\Mott}$ for the decay direction (1,1,1).}
  \label{fig:beta_distribution_111} 
\end{figure}

While Table \ref{table:2} gives us an idea of the decay rates at specific single directions, in reality the wavefunctions may not be perfectly propagating along a given direction. The non-perpendicular incident components can be taken into account by simply allowing $C_{2}$ and $C_{3}$ (defined in Eq.(5)) to vary, resulting in a distribution of $\beta (C_{2},C_{3})$ for a certain decay direction $\hat{k}_{\perp}$. This means we assume the tunnelling barrier system has translational symmetry in the plane parallel to the interface, so that the transmission conserves $\vec{k}_{\parallel}$. At $E_{B}^{\Mott}$, we have obtained the $\beta (C_{2},C_{3})$ as shown in Fig.\ref{fig:beta_distribution_100}, Fig.\ref{fig:beta_distribution_110}, and Fig.\ref{fig:beta_distribution_111}. 

\begin{figure}[H]
  \centering
  \begin{tabular}{@{}c@{}}
    \includegraphics[width=6cm,height=2cm]{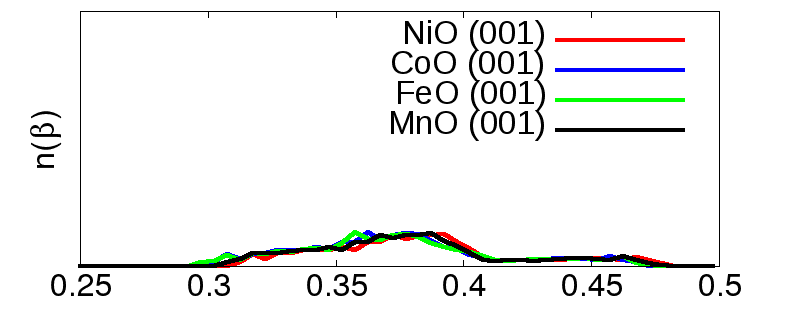}
  \end{tabular}
  \begin{tabular}{@{}c@{}}
    \includegraphics[width=6cm,height=2cm]{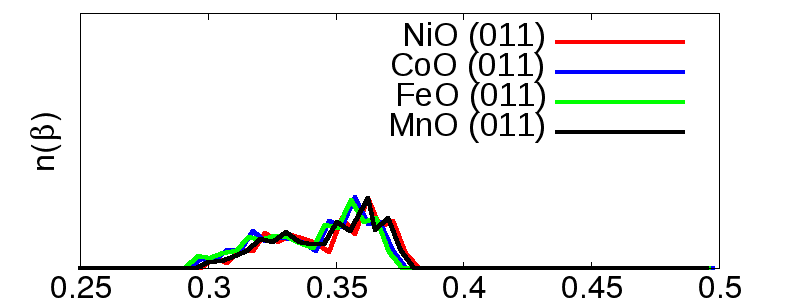}
  \end{tabular}
  \begin{tabular}{@{}c@{}}
    \includegraphics[width=6cm,height=2cm]{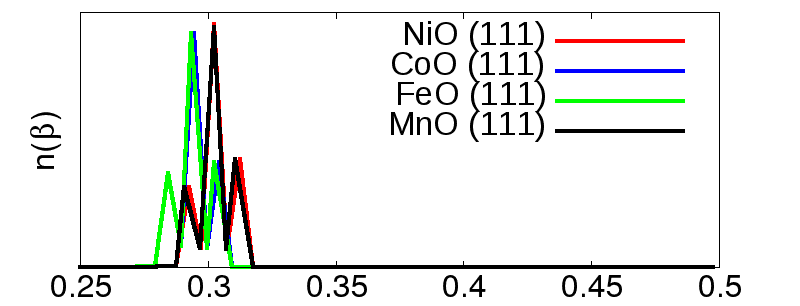}
  \end{tabular}
  \caption{$\beta$-resolved density of state $n(\beta)$ for the three directions: (0,0,1) upper, (0,1,1) middle, (1,1,1) down. The $\beta$ is measured at $E_{B}^{\Mott}$. The $y$-axis scale is same for the three plots. }
  \label{fig:beta_density}    
\end{figure}

At first glance, the symmetry of these distributions is consistent with the crystal symmetry, which makes sense because single-site DMFT assumes a completely localized self-energy. The ranges of values of $\beta$ are quite different in different directions. The observation becomes clearer when one introduces the $\beta$-resolved density of states $n(\beta)$. \cite{PRL_beta_density} Ideally $n(\beta) \, d\beta$ would be the number of "states" with $\beta$ values in the infinitesimal interval $(\beta,\beta+d\beta)$. Numerically, we calculated $\beta$ on the grid of $C_{2} \times C_{3} = 80 \times 80 $ points within [0,1]x[0,1], and applied linear interpolation to make the grid denser. The resulting density is shown in Fig.\ref{fig:beta_density}. It is clear that the decay rates cluster around 0.3 for the (1,1,1) direction, and extend to larger and larger values in (0,1,1) and (0,0,1) directions. 


\begin{figure}[H] 
  
  \begin{subfigure}[b]{0.5\linewidth} 
    \centering
    \includegraphics[width=1.0\columnwidth]{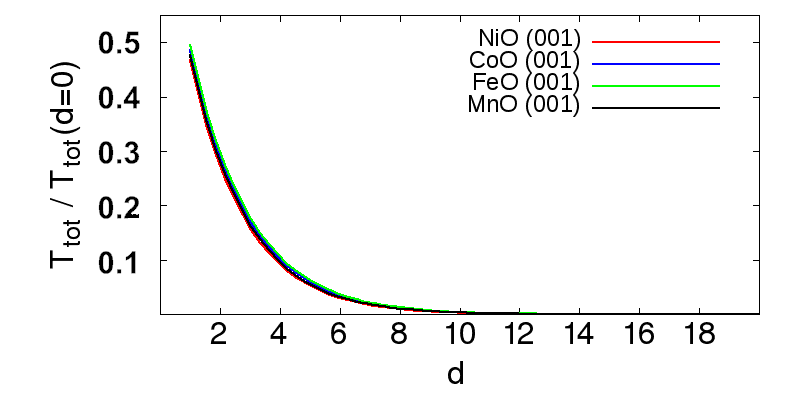} 
    \caption{(0,0,1)} 
  \end{subfigure}
  \begin{subfigure}[b]{0.5\linewidth}
    \centering
    \includegraphics[width=1.0\columnwidth]{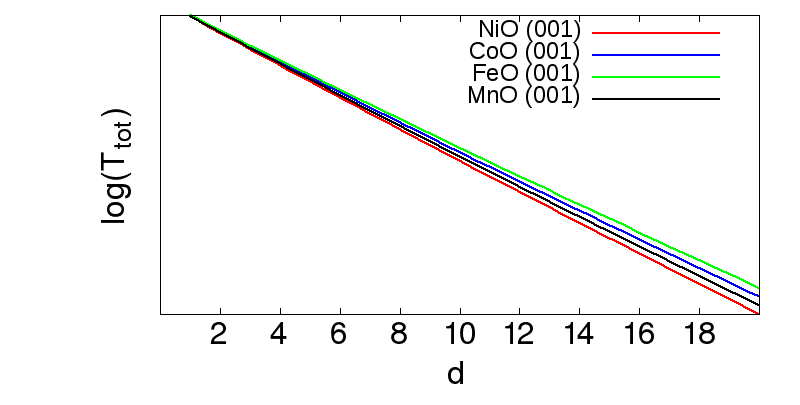} 
    \caption{(0,0,1), log scale} 
  \end{subfigure} 
  
  \begin{subfigure}[b]{0.5\linewidth}
    \centering
    \includegraphics[width=1.0\columnwidth]{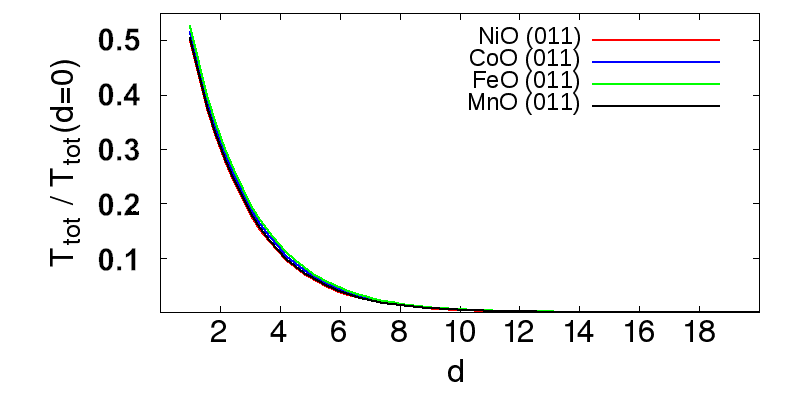} 
    \caption{(0,1,1)} 
  \end{subfigure}
  \begin{subfigure}[b]{0.5\linewidth}
    \centering
    \includegraphics[width=1.0\columnwidth]{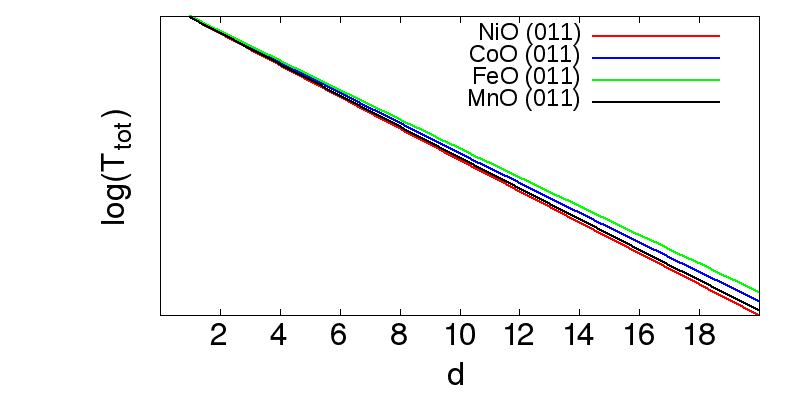} 
    \caption{(0,1,1), log scale} 
  \end{subfigure}
  
  \begin{subfigure}[b]{0.5\linewidth}
    \centering
    \includegraphics[width=1.0\columnwidth]{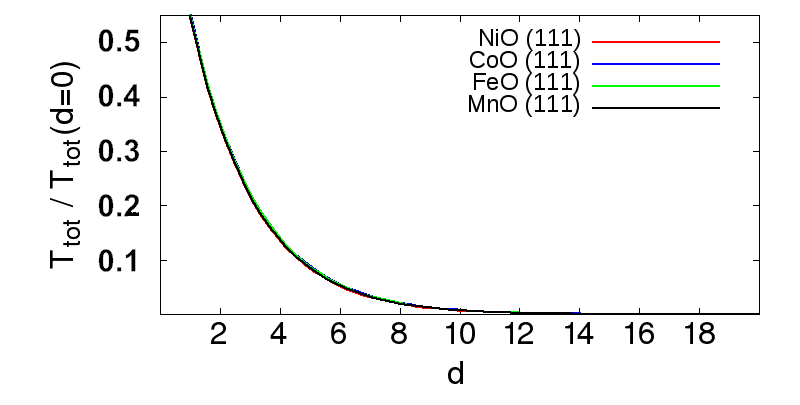} 
    \caption{(1,1,1)} 
  \end{subfigure}
  \begin{subfigure}[b]{0.5\linewidth}
    \centering
    \includegraphics[width=1.0\columnwidth]{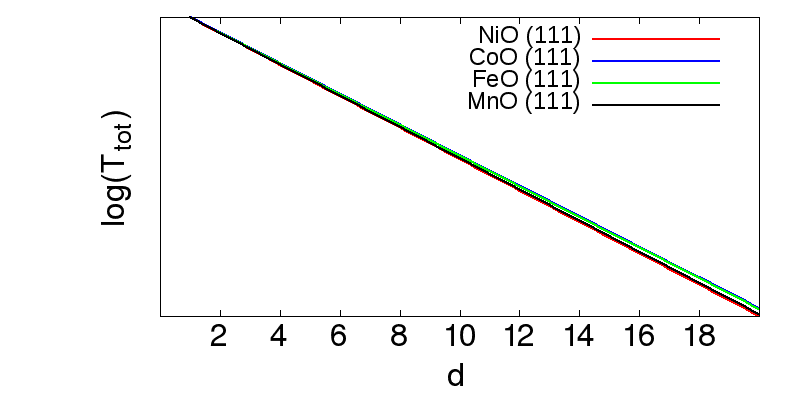} 
    \caption{(1,1,1), log scale} 
  \end{subfigure}
  \caption{Relative transmission probability on linear and log scale for the
directions: (0,0,1), upper two plots; (0,1,1), middle two plots; (1,1,1), lower
two plots; }
  \label{fig:beta_transmission1} 
\end{figure}

In general, the probability for wavefunction transmission through the tunnel junction can be written as the product of the probability for transmission across each of the interfaces times a factor that describes the exponential decay of the electron probability within the junction material \cite{0953-8984-15-41-R01,PhysRevB.80.035124}.  
\begin{equation} 
T_{B}(\vec{k}_{\parallel}) = T_{L}(\vec{k}_{\parallel}) \cdot T_{R}(\vec{k}_{\parallel}) \cdot T^{\evan}_{\tot}(d)
\end{equation} 
Here $T_{B}$ is the transmission probability through the junction. $T_{L}$ and $T_{R}$ are the probabilities for an electron to be transmitted across the left and right electrode barrier interfaces respectively, and $T^{\evan}_{\tot}$ is the evanescent channel contribution to the conductance. We use a simplified model to estimate the evanescent channel transmission probability $T^{\evan}_{\tot}$. The transmission probability for each $\vec{k}_{\parallel}$ takes the form: $T^{\evan}_{\vec{k}_{\parallel}}(d)=T_{0} \cdot exp(-\beta_{\vec{k}_{\parallel}} \cdot d)$, where $d$ is the thickness (number of layers) of the tunneling barrier and $\beta_{\vec{k}_{\parallel}}$ is the distribution of $\beta(C_{2},C_{3})$ obtained in Fig.\ref{fig:beta_distribution_100}, Fig.\ref{fig:beta_distribution_110}, and Fig.\ref{fig:beta_distribution_111}. The total evanescent transmission probability for the direction $\hat{k}_{\perp}$, at a given energy level, is given by:
\begin{equation} 
T^{\evan}_{\tot}(d) = \frac{1}{N_{\vec{k}_{\parallel}}} \sum_{\vec{k}_{\parallel}} T^{\evan}_{\vec{k}_{\parallel}}(d)
\end{equation} 
We calculate the relative evanescent transmission probability $T^{\evan}_{\tot}(d)/T^{\evan}_{\tot}(d=0)$ at $E_{B}^{\Mott}$, as shown in Fig.\ref{fig:beta_transmission1}. The common feature of exponential decay is clear, and the relative transmission becomes very small after about 10 layers of the unit cell, for all four materials. The larger values of $\beta$ in the (0,0,1) direction results in a slightly faster decay than for the other two directions. The differences in $T^{\evan}_{\tot}(d)$ between the three directions is overall not significant after averaging $\vec{k}_{\parallel}$. 

\section{\label{sec:4}CONCLUSION}

In summary, we have performed DFT plus single-site DMFT calculation for the four transition metal monoxides, NiO, CoO, FeO and MnO, in the non-spin-polarized phase, and have studied the complex band structures of them by including the DMFT self-energy in the Green's function. The $d$ and $p$ orbitals are included in the DMFT calculation, and the TM $4s$ orbital is included in the Green's function for complex band analysis. Both the $dp$ and $d$-$dp$ models have been considered, and the corresponding screened Coulomb interaction parameters are calculated from first-principles using the cRPA method. The resulting spectral functions of the $d$-$dp$ model present a clear band gap in general agreement with experiments, improving upon the gapless DFT ground state calculation. 

By using the full Green's function that includes $s$, $p$, $d$ orbitals and the DMFT self-energy, we have observed the complex band structures of these Mott insulators. Motivated by the tunnel junction application of these materials, we analyzed the pinning position of the Fermi level by applying the charge neutrality condition. In the complex domain, the gap states transitioning from valence-band character to conduction-band character displays a jump in $dE/dk$, rather than the continuous transition seen in traditional semiconductors. The branch point is found to be at the jump point, and is very close to conduction band minimum. The TM $4s$ band is carefully studied, and we argue that the 4$s$ band is not responsible for determining the Fermi level pinning position, which is supported by experiment on NiO tunnel junctions. The calculated results are in consistent with experimental observations that NiO has large band gap and small Schottky barrier height. The transmission decay parameter, $\beta$, has been calculated for different directions in $k$-space to give us insight into the evanescent channel contribution to the conductance. We have investigated $\beta$ in detail at the Mott branch point within the correlation window. We found that the $\beta$ parameter has very different values and distributions for different directions. When the decay direction and the incident direction are not the same, $\beta$ generally becomes larger and forms non-trivial patterns depending on the relative direction. 

The CBS analysis relies on the DMFT self-energy. Our DFT+DMFT calculation based on the modified ELK code and DCA++ code is a single shot implementation, not a fully charge density self-consistent one. The current work assumed the TM 4$s$ band is not significantly shifted in a CSC treatment, which is reasonable based on existing studies. We noticed the existing CSC DFT+DMFT calculations of the same materials have shown the necessity of updating charge density when structural changes due to pressure and strain are closely tied to electronic transitions, which raises interesting direction to motivate our future work. 

In conclusion, the presented work carries out a fully \textit{ab initio} study of the charge neutrality level and wavefunction decay rate in the evanescent channel of late transition metal monoxides in their PM phase under ambient condition, by using the combination of CBS method and DFT+DMFT calculation. The DFT+DMFT band gap calculations, along with the cRPA calculations of Coulomb $U$ matrices, are done in a standard way. The main physical features are captured. The newly observed feature of the CBS and location of branch point of these Mott insulators are different from band insulators. In addition, the TM 4$s$ band of NiO is found to have different feature in the complex domain than that of the other three materials. The barrier height, values of decay rate and its direction dependence can all be obtained from first principle. This numerical study could be useful when Mott insulating materials are used for tunnel junction applications. The approach could be applied to more complicated structures or lower dimensional cases, as long as the application of singe-site DFT+DMFT is justified. 


\section{\label{sec:5}Acknowledgements}

This work was supported by the US Department of Energy (DOE), Office of Basic Energy Sciences (BES), under Contract No.~DE-FG02-02ER45995. The computation was done using the utilities of the National Energy Research Scientific Computing Center (NERSC).


\section{Appendix A}\label{appendix_a}

The on-site Coulomb interaction U-matrix is required input for the impurity problem. One reliable way to calculate these parameters from first principles is the constrained Random Phase Approximation(cRPA) method, which has been well described in the literatures \cite{PhysRevB_frequ_depend_U,U_from_cRPA}. Here we first summarize the original idea. Then we explain our implementation based on the density response function, and list our results of U-matrices for the two models and the four materials. 

The cRPA calculation is based on a DFT ground state calculation that includes many empty bands. One aims to get an estimation of the screened Coulomb interaction for the selected bands of interest, or an energy window. For this purpose, the particle-hole polarization between all possible pairs of occupied state and unoccupied state are taken into account. Within the RPA, the particle-hole polarization is calculated as \cite{TDDFT_density_response_PRL_76_1212}:
\begin{multline}
P_{\tot}(r,r\ensuremath{'};\omega)=\sum_{i}^{occ.}\sum_{j}^{unocc.}\psi_{i}^{*}(r)\psi_{j}(r\ensuremath{'})\psi_{j}^{*}(r)\psi_{i}(r\ensuremath{'})\times \\ 
(\frac{1}{\omega-\varepsilon_{j}+\varepsilon_{i}+i\delta}+\frac{1}{\omega+\varepsilon_{j}-\varepsilon_{i}-i\delta})
\end{multline}
where $\psi_{i}$ and $\varepsilon_{i}$ are the eigenfunctions and eigenenergies of the one-particle Hamiltonian in DFT. 

The selected bands of interest are often around the Fermi level, and have a particular orbital character, for example $d$-like in our case. Following the convention in the literatures, we label the bands of interest or energy window as the $d$-space. If both the occupied state and the unoccupied state are within the $d$-space, then the polarization contributes to $P_{d}(r,r\ensuremath{'};\omega)$. All the other pairs of occupied and unoccupied states contribute to $P_{r}$, where $r$ stands for the rest of the bands. Thus, the total polarization is divided into two parts: $P_{\tot} = P_{d} + P_{r}$. $P_{r}$ is related to the partially screened Coulomb interaction \cite{PhysRevB_frequ_depend_U}: 
\begin{equation} 
W_{r}(\omega)=[1-\nu \cdot P_{r}(\omega)]^{-1}\cdot\nu 
\end{equation} 
In the above equation, $\nu$ is the bare Coulomb interaction. It's obvious that the total polarization, $P_{\tot}$, screens the bare Coulomb interaction, $\nu$, to give the fully screened interaction $W$. With the same logic, $P_{d}$ screens $W_{r}$ to give the fully screened interaction $W$. Thus, $W_{r}$ is identified as the on-site Coulomb interaction for the bands of interest, $i.e.$ $U(\omega) \equiv W_{r}(\omega)$, which has included the screening effect from the realistic environment of the material. 

Our cRPA calculation is based on the partial Kohn-Sham susceptibility \cite{TDDFT_density_response_PRL_76_1212,Anton_IEEE_paper_2010}: 
\begin{equation}
\chi^{KS}_{r}(r,r\ensuremath{'};\omega)=\sum_{i,j \not\subset \textit{C.S.}}\frac{(f_{i}-f_{j})\psi_{i}^{*}(r)\psi_{j}(r\ensuremath{'})\psi_{j}^{*}(r)\psi_{i}(r\ensuremath{'})}{\omega-\varepsilon_{j}+\varepsilon_{i} +i\delta}
\end{equation}
where $f_{i}$ and $\varepsilon_{i}$ are the occupancy and energy of the eigenstate $\psi_{i}$. The summation over band indices runs over all bands excluding the cases where both $i$ and $j$ are inside the correlation subspace (\textit{C.S.} under the summation sign means correlation subspace). In general, the density response function $\chi(r,r\ensuremath{'};\omega)$ is related to the Kohn-Sham susceptibility by the following integral equation \cite{TDDFT_density_response_PRL_76_1212}: 
\begin{multline}
\chi(r,r\ensuremath{'};\omega)=\chi^{KS}(r,r\ensuremath{'};\omega) + \iint dr_{1}dr_{2} \ \chi^{KS}(r,r_{1};\omega) \\
\times \left( \frac{1}{|r_{1}-r_{2}|}+f^{xc}(r_{1},r_{2};\omega)\right) \times \chi(r_{2},r\ensuremath{'};\omega)
\end{multline}
where the $f^{xc}$ is the functional derivative of the exchange-correlation potential with respect to the charge density, which is often neglected in the random phase approximation. Since we are considering periodic crystal structures, the integral equation is often written in the Fourier-transformed form where $\chi(r,r\ensuremath{'};\omega)$ becomes $\chi(q,q\ensuremath{'};\omega)$ with $q$ and $q'$ the reciprocal lattice vectors. Due to the invariance of the real space response function with respect to a shift by a lattice vector $R$: $\chi(r+R,r\ensuremath{'}+R;\omega)=\chi(r,r\ensuremath{'};\omega)$, the $\chi(q,q\ensuremath{'};\omega)$ is only nonezero when $q$ and $q'$ differ by a reciprocal lattice vector $G$. One can replace $q$ by $q+G$, replace $q'$ by $q+G'$, and restrict $q$ to be always within the first Brillouin zone. Thus the Fourier-transformed integral equation has the form: 
\begin{multline}
\chi_{GG'}(q,\omega)=\chi^{KS}_{GG'}(q,\omega) + \sum_{G_{1}G_{2}} \chi^{KS}_{GG_{1}}(q,\omega)\times \\ 
\left ( v_{G_{1}+q}\delta_{G_{1}G_{2}} + f^{xc}_{G_{1}G_{2}}(q,\omega) \right ) \times \chi_{G_{2}G'}(q,\omega)
\end{multline}
where $v_{G+q}=4\pi /|G+q|^{2}$ is the expansion coefficient of the bare Coulomb interaction. By using the partial Kohn-Sham susceptibility $\chi^{KS}_{r,GG'}(q,\omega)$ (Fourier transform of Eq.(6)) and neglecting the $f^{xc}$ term in Eq.(7), we reach the equation for the partial RPA density response function $\chi_{r}^{RPA}(q,\omega)$: 
\begin{multline}
\chi_{r,GG'}^{RPA}(q,\omega)=\chi^{KS}_{r,GG'}(q,\omega) + \\ 
\sum_{G_{1}} v_{G_{1}+q} \times \chi^{KS}_{r,GG_{1}}(q,\omega) \times \chi_{r,G_{1}G'}^{RPA}(q,\omega)
\end{multline}
Note that the subscript $r$ in Eq.(9) stands for the $rest$ of the bands, as same as in Eq.(5). Eq.(9) is first solved for $\chi_{r}^{RPA}(q,\omega)$ in the calculation. The rest calculation is based on the linear respone theory \cite{Wr_related_to_invDielecFunc}, where the partially screened Coulomb interaction $W_{r}$ is related to inverse dielectric function $\varepsilon^{-1}$ and bare Coulomb interaction $\nu$: $W_{r}(r_{1},r_{2};\omega)=\int dr \varepsilon^{-1}(r_{1},r;\omega)\nu(r,r_{2})$. The inverse dielectric function $\varepsilon^{-1}$ is determined by $\varepsilon^{-1}(r_{1},r;\omega)=1+\nu\cdot\chi_{r}^{RPA}(r_{1},r;\omega)$. Finally the frequency-dependent screened Coulomb interaction is computed from the partial RPA density response function and the bare Coulomb interaction:
\begin{multline}
W_{r,GG'}(q,\omega)=v_{G+q}\delta_{GG'} + v_{G+q} \cdot \chi_{r,GG'}^{RPA}(q,\omega) \cdot v_{G'+q}
\end{multline}
The above calculations have been implemented in the Exciting-Plus code (a modified version of ELK code) \cite{ELK_code,Anton_IEEE_paper_2010}, which we used for the U matrix calculations. The Wannier orbitals are constructed by projection to preserve symmetry, and no spatial localization procedure is applied. In addition to the DFT calculation described in Sec.II.A, 100 empty bands are included for the cRPA calculation. 


NiO, $dp$ model: 
\[ 
U_{mm^{'}}^{\sigma\overline{\sigma}} =
\left( \begin{array}{ccccc}
9.69 & 8.03 & 7.97 & 8.03 & 9.10 \\
8.03 & 9.69 & 8.82 & 8.03 & 8.25 \\
7.97 & 8.82 & 10.27 & 8.82 & 8.19 \\
8.03 & 8.03 & 8.82 & 9.69 & 8.25 \\
9.10 & 8.25 & 8.19 & 8.25 & 10.27 \end{array} \right) 
\] 
\[ 
U_{mm^{'}}^{\sigma\sigma} =
\left( \begin{array}{ccccc}
0.00 & 7.21 & 6.97 & 7.21 & 8.66 \\
7.21 & 0.00 & 8.24 & 7.21 & 7.39 \\
6.97 & 8.24 & 0.00 & 8.24 & 7.15 \\
7.21 & 7.21 & 8.24 & 0.00 & 7.39 \\
8.66 & 7.39 & 7.15 & 7.39 & 0.00 \end{array} \right) 
\] 
NiO, $d$-$dp$ model: 
\[ 
U_{mm^{'}}^{\sigma\overline{\sigma}} =
\left( \begin{array}{ccccc}
7.29 & 5.67 & 5.42 & 5.67 & 6.48 \\
5.67 & 7.29 & 6.21 & 5.67 & 5.69 \\
5.42 & 6.21 & 7.38 & 6.21 & 5.44 \\
5.67 & 5.67 & 6.21 & 7.29 & 5.69 \\
6.48 & 5.69 & 5.44 & 5.69 & 7.38 \end{array} \right) 
\] 
\[ 
U_{mm^{'}}^{\sigma\sigma} =
\left( \begin{array}{ccccc}
0.00 & 4.85 & 4.44 & 4.85 & 6.04 \\
4.85 & 0.00 & 5.64 & 4.85 & 4.84 \\
4.44 & 5.64 & 0.00 & 5.64 & 4.47 \\
4.85 & 4.85 & 5.64 & 0.00 & 4.84 \\
6.04 & 4.84 & 4.47 & 4.84 & 0.00 \end{array} \right) 
\] 
CoO, $dp$ model: 
\[ 
U_{mm^{'}}^{\sigma\overline{\sigma}} =
\left( \begin{array}{ccccc}
9.27 & 7.70 & 7.69 & 7.70 & 8.73 \\
7.70 & 9.27 & 8.47 & 7.70 & 7.95 \\
7.69 & 8.47 & 9.92 & 8.47 & 7.93 \\
7.70 & 7.70 & 8.47 & 9.27 & 7.95 \\
8.73 & 7.95 & 7.93 & 7.95 & 9.92 \end{array} \right) 
\] 
\[ 
U_{mm^{'}}^{\sigma\sigma} =
\left( \begin{array}{ccccc}
0.00 & 6.92 & 6.73 & 6.92 & 8.30 \\
6.92 & 0.00 & 7.91 & 6.92 & 7.12 \\
6.73 & 7.91 & 0.00 & 7.91 & 6.93 \\
6.92 & 6.92 & 7.91 & 0.00 & 7.12 \\
8.30 & 7.12 & 6.93 & 7.12 & 0.00 \end{array} \right) 
\] 
CoO, $d$-$dp$ model: 
\[ 
U_{mm^{'}}^{\sigma\overline{\sigma}} =
\left( \begin{array}{ccccc}
6.47 & 4.94 & 4.71 & 4.94 & 5.68 \\
4.94 & 6.47 & 5.43 & 4.94 & 4.96 \\
4.71 & 5.44 & 6.57 & 5.44 & 4.74 \\
4.94 & 4.94 & 5.44 & 6.47 & 4.96 \\
5.68 & 4.96 & 4.74 & 4.96 & 6.57 \end{array} \right) 
\] 
\[ 
U_{mm^{'}}^{\sigma\sigma} =
\left( \begin{array}{ccccc}
0.00 & 4.16 & 3.78 & 4.16 & 5.25 \\
4.16 & 0.00 & 4.89 & 4.16 & 4.15 \\
3.78 & 4.89 & 0.00 & 4.89 & 3.82 \\
4.16 & 4.16 & 4.89 & 0.00 & 4.15 \\
5.25 & 4.15 & 3.82 & 4.15 & 0.00 \end{array} \right) 
\] 
FeO, $dp$ model: 
\[ 
U_{mm^{'}}^{\sigma\overline{\sigma}} =
\left( \begin{array}{ccccc}
8.91 & 7.49 & 7.53 & 7.49 & 8.40 \\
7.49 & 8.91 & 8.18 & 7.49 & 7.75 \\
7.53 & 8.18 & 9.59 & 8.18 & 7.79 \\
7.49 & 7.49 & 8.18 & 8.91 & 7.75 \\
8.40 & 7.75 & 7.79 & 7.75 & 9.59 \end{array} \right) 
\] 
\[ 
U_{mm^{'}}^{\sigma\sigma} =
\left( \begin{array}{ccccc}
0.00 & 6.76 & 6.64 & 6.76 & 7.98 \\
6.76 & 0.00 & 7.65 & 6.76 & 6.98 \\
6.64 & 7.65 & 0.00 & 7.65 & 6.89 \\
6.76 & 6.76 & 7.65 & 0.00 & 6.98 \\
7.98 & 6.98 & 6.89 & 6.98 & 0.00 \end{array} \right) 
\] 
FeO, $d$-$dp$ model: 
\[ 
U_{mm^{'}}^{\sigma\overline{\sigma}} =
\left( \begin{array}{ccccc}
5.89 & 4.50 & 4.33 & 4.50 & 5.13 \\
4.50 & 5.89 & 4.93 & 4.50 & 4.53 \\
4.33 & 4.93 & 6.02 & 4.93 & 4.35 \\
4.50 & 4.50 & 4.93 & 5.89 & 4.53 \\
5.13 & 4.53 & 4.35 & 4.53 & 6.02 \end{array} \right) 
\] 
\[ 
U_{mm^{'}}^{\sigma\sigma} =
\left( \begin{array}{ccccc}
0.00 & 3.78 & 3.47 & 3.78 & 4.71 \\
3.78 & 0.00 & 4.40 & 3.78 & 3.78 \\
3.47 & 4.40 & 0.00 & 4.40 & 3.52 \\
3.78 & 3.78 & 4.40 & 0.00 & 3.78 \\
4.71 & 3.78 & 3.52 & 3.78 & 0.00 \end{array} \right) 
\] 
MnO, $dp$ model: 
\[ 
U_{mm^{'}}^{\sigma\overline{\sigma}} =
\left( \begin{array}{ccccc}
8.68 & 7.34 & 7.39 & 7.34 & 8.18 \\
7.34 & 8.68 & 7.99 & 7.34 & 7.59 \\
7.39 & 7.99 & 9.34 & 7.99 & 7.65 \\
7.34 & 7.34 & 7.99 & 8.68 & 7.59 \\
8.18 & 7.59 & 7.65 & 7.59 & 9.34 \end{array} \right) 
\] 
\[ 
U_{mm^{'}}^{\sigma\sigma} =
\left( \begin{array}{ccccc}
0.00 & 6.65 & 6.56 & 6.65 & 7.78 \\
6.65 & 0.00 & 7.47 & 6.65 & 6.87 \\
6.56 & 7.47 & 0.00 & 7.47 & 6.80 \\
6.65 & 6.65 & 7.47 & 0.00 & 6.87 \\
7.78 & 6.87 & 6.80 & 6.87 & 0.00 \end{array} \right) 
\] 
MnO, $d$-$dp$ model: 
\[ 
U_{mm^{'}}^{\sigma\overline{\sigma}} =
\left( \begin{array}{ccccc}
5.59 & 4.28 & 4.13 & 4.28 & 4.86 \\
4.28 & 5.59 & 4.68 & 4.28 & 4.31 \\
4.13 & 4.68 & 5.73 & 4.68 & 4.17 \\
4.28 & 4.28 & 4.68 & 5.59 & 4.31 \\
4.86 & 4.31 & 4.17 & 4.31 & 5.73 \end{array} \right)
\] 
\[ 
U_{mm^{'}}^{\sigma\sigma} =
\left( \begin{array}{ccccc}
0.00 & 3.60 & 3.33 & 3.60 & 4.46 \\
3.60 & 0.00 & 4.18 & 3.60 & 3.61 \\
3.33 & 4.18 & 0.00 & 4.18 & 3.38 \\
3.60 & 3.60 & 4.18 & 0.00 & 3.61 \\
4.46 & 3.61 & 3.38 & 3.61 & 0.00 \end{array} \right)
\] 


The cRPA calculation directly provides the intra-, inter-orbital and exchange interaction parameters. We derived single value intra-, inter-orbital and exchange interaction parameters $U$, $U\ensuremath{'}$ and $J$ from averaging the matrix elements, in order to compare with the existing calculations. By definition, the diagonal elements of $U_{mm\ensuremath{'}}^{\sigma\overline{\sigma}}$ are the intra-orbital interactions, and the average value is: $U=\frac{1}{5}TrU_{mm\ensuremath{'}}^{\sigma\overline{\sigma}}$. The inter-orbital interactions are the off-diagonal elements of $U_{mm\ensuremath{'}}^{\sigma\overline{\sigma}}$, thus $U\ensuremath{'}= \frac{1}{20}\sum_{m\neq m\ensuremath{'}}U_{mm\ensuremath{'}}^{\sigma\overline{\sigma}}$. The inter-orbital exchange is the off-diagonal elements of $U_{mm\ensuremath{'}}^{\sigma\overline{\sigma}}-U_{mm\ensuremath{'}}^{\sigma\sigma}$, so $J = \frac{1}{20}\sum_{m \neq m\ensuremath{'}} ( U_{mm\ensuremath{'}}^{\sigma\overline{\sigma}}-U_{mm\ensuremath{'}}^{\sigma\sigma} ) $. The values of $U$, $U\ensuremath{'}$ and $J$ for the two models are summarized in Table \ref{table:3}. We found general agreement with other cRPA calculation of the same materials in the literature\cite{U_from_MLWF_cRPA_NiOCoOFeOMnO_2013}. 

    \begin{table}[ht]
    \fontsize{10}{14}\selectfont
    \centering
    \begin{tabular}{ | >{\centering}m{1.8cm} | >{\centering}m{1.5cm} | >{\centering}m{1.5cm} | >{\centering}m{1.5cm} | c |}
    \hline
       (eV)             & NiO  & CoO  & FeO  & MnO   \\ \hline 
    $U$, $dp$   & 9.92(10.3) & 9.53(9.8) & 9.09(9.5) & 8.95(9.2)  \\  
    $U$, $d$-$dp$       & 7.33(7.6) & 6.51(6.8) & 5.94(6.3) & 5.65(6.1)  \\ \hline 
    $U\ensuremath{'}$, $dp$       & 8.35(8.6) & 8.03(8.1) & 7.71(7.9) & 7.64(7.7)  \\ 
    $U\ensuremath{'}$, $d$-$dp$   & 5.81(5.9) & 5.07(5.2) & 4.62(4.8) & 4.40(4.6)  \\ \hline
    $J$, $dp$           & 0.78(0.9) & 0.75(0.8) & 0.70(0.8) & 0.66(0.7)  \\ 
    $J$, $d$-$dp$       & 0.77(0.9) & 0.73(0.8) & 0.68(0.8) & 0.64(0.7)  \\ \hline
    \end{tabular}
    \caption{The values of $U$, $U\ensuremath{'}$ and $J$ deduced from the cRPA calculation. Values in parenthesis are from Ref.\cite{U_from_MLWF_cRPA_NiOCoOFeOMnO_2013} for the exactly same model construction using a different code based on Maximum Localized Wannier Functions.}
    \label{table:3}
    \end{table}


\clearpage

\vspace{10mm}

\bibliography{ms}

\end {document}